\documentclass[aps,prd,twocolumn,groupedaddress,showpacs,floatfix]{revtex4}

\usepackage{graphicx}
\usepackage{dcolumn}

\def\slash#1{\mkern-1.5mu\raise0.4pt\hbox{$\not$}\mkern1.2mu #1\mkern 0.7mu}

\begin{document}

\title{Chiral violations in domain-wall QCD 
        \\ from one-loop perturbation theory at finite $N_s$}

\author{Stefano Capitani}
\email[]{stefano.capitani@uni-graz.at}
\affiliation{Institut f\"ur Physik, FB Theoretische Physik \\
Universit\"at Graz, A-8010 Graz, Austria}

\begin{abstract}
We present perturbative calculations made with domain-wall fermions which
possess a finite number of points $N_s$ in the extra fifth dimension. We have 
derived the required propagator functions, investigated the one-loop properties
of quark amplitudes at finite $N_s$ and evaluated three quantities that can 
provide insights on chirality-breaking effects from the perturbative side.

First we have computed the residual mass for various choices of $N_s$
and of the domain-wall height $M$. We have found that this radiatively induced 
mass approaches zero reasonably fast with the extent of the fifth dimension, 
depending on $M$ and on a lesser extent on the coupling $g_0$. We have also 
computed the differences of the renormalization constants of the vector and 
axial-vector currents and of the scalar and pseudoscalar densities. 
Finally we have calculated the chirally-forbidden mixing (which at finite $N_s$
is suppressed only partially) of an operator which describes the 
lowest moment of the $g_2$ structure function. In general we see that at 
$M=1.8$, where simulations are usually performed, values of $N_s=20$ or larger 
would be desirable in order for chiral violations to be negligible.

The quantities that we have studied turn out to lose gauge invariance when 
$N_s$ is not infinite. We have also found that anomalous dimensions of 
operators at finite $N_s$ generally depend on $N_s$ and $M$. In particular, 
the vector and axial-vector currents have in general a nonzero anomalous 
dimension at finite $N_s$.
\end{abstract}


\maketitle

\section{Introduction}

Domain-wall fermions \cite{Kaplan:1992bt,Shamir:1993zy,Furman:1994ky}
constitute one of the known solutions of the Ginsparg-Wilson relation 
\cite{Ginsparg:1981bj} and are hence invariant with respect to chiral symmetry 
transformations even away from the continuum limit, for nonvanishing values 
of the lattice spacing $a$ \cite{Luscher:1998pq}, while avoiding 
at the same time unpleasant effects like doublers and nonanaliticities. 
The massless chiral mode generated by the domain wall survives 1-loop 
renormalization and the theory has been proven to be renormalizable at this 
order \cite{Yamada:1997bj,Yamada:1997hc}. The possibility arises of the 
construction of chiral gauge theories at finite $a$ \cite{Kikukawa:2001mw}. 
Moreover, with this kind of fermions the leading discretization errors 
are reduced to $O(a^2)$ when chiral symmetry is exact. 

A certain amount of chiral symmetry breaking arises however in Monte Carlo 
simulations of domain-wall fermions, because they must be performed using 
lattices which have a finite number of points, $N_s$, in the fifth 
dimension. It is only in the theoretical limit in which the extension of the 
fifth dimension becomes infinite that the chiral modes (which are exponentially
confined on the two opposite walls) can fully decouple from each other, 
yielding an exact chiral symmetry. The chiral modes acquire some mass
if the distance between the two walls is not infinite, and to study the
magnitude of these chirality-violating effects is one of the main objectives 
of the present work.
With the computer speeds presently available it is unfortunately not yet 
possible to obtain significant physics from simulations performed at a large 
$N_s$, where these chiral violations would be numerically negligible. 
After the first pioneering Monte Carlo implementations of domain-wall fermions 
\cite{Blum:1996jf,Blum:1997mz} and subsequent advances reported in 
\cite{Aoki:2000pc,Chen:2000zu,Blum:2000kn,Blum:2001xb,Noaki:2001un} and 
\cite{Aoki:2002vt,Aoki:2004ht}, the most recent simulations, which have 
considered phenomenological quantities as diverse as weak matrix elements, 
structure functions and heavy-light meson spectroscopy, have been mostly 
performed using lattices with only $N_s=12$ or $16$ 
(for a selection of the latest results see for example 
\cite{Orginos:2005uy,Aoki:2005ga,Negele:2005za,Yamada:2005dv,Lin:2005gh,Edwards:2005kw,Dawson:2005zv,Noaki:2005zw,Edwards:2005ym,Ohta:2005cn,Hashimoto:2005re,Berruto:2005hg,Berruto:2005sy,Lin:2006kg}). 

The experiences gained in these recent investigations seem to indicate that 
for extents of the extra fifth dimension as small as $N_s=16$ the 
chirality-breaking effects \cite{Vranas:1997da,Vranas:1997ib,Jung:2000fh,Gadiyak:2002ig,Blum:2001xb,Edwards:2005an} are still under control. The residual mass 
$m_{res}$ for typical lattice spacings of about $a^{-1}=1.5-2$ GeV generally 
turns out to be of $O(10^{-3})$ or $O(10^{-2})$, depending on whether 
simulations are carried out in quenched or full QCD, and on which type of gauge
action is used \cite{Aoki:2002vt,Aoki:2004ht,Lin:2005gh}. In particular, when 
using quenched QCD instead of the full theory, or renormalization group 
improved gauge actions (like Iwasaki and DBW2) instead of the simple plaquette 
action, the residual mass becomes substantially reduced. 
This is related to the fact that, for a fixed choice of $a$, the values of 
$\beta = 6/g_0^2$ are larger in those cases, and the gauge fields smoother. 
The residual mass is then in general not very small (expecially in the full QCD
case), and it can at times become comparable to the input light sea quark 
masses. However, if the exponential suppression of chirality-breaking effects 
takes place rather quickly with the length of the fifth dimension, then 
increasing $N_s$ a little further could already be sufficient to obtain 
at last almost negligible chirality-breaking effects. 

The extent to which chiral symmetry is broken in lattices with a small $N_s$ 
is thus one of the most important issues which need to be understood in present
domain-wall simulations. To the extent that one-loop calculations can provide 
clues to the true behavior of the truncated domain-wall theory at finite $N_s$,
we believe that it is useful and interesting to investigate chirality-breaking 
effects also from the point of view of perturbation theory, complementing 
nonperturbative investigations of such effects. 
Towards this end we present here the results of some selected one-loop 
calculations which we have carried out using the Feynman rules which exactly  
correspond to the theory at finite $N_s$. This is at variance with past 
domain-wall perturbative calculations where, in place of the exact quark 
propagators, their asymptotic expressions for large $N_s$ were instead used 
\cite{Aoki:1997xg,Aoki:1998vv,Blum:1999xi,Capitani:2005vb}. 
The purpose of this work is to calculate with the exact Feynman rules the 
deviations from the $N_s=\infty$ results in the case where $N_s$ is limited 
to small values of $O(10)$, that is for situations which roughly correspond 
to present simulations. We use the plaquette gauge action, and we can obtain 
some estimates of the amount of perturbative chiral violations by focusing 
on quantities like the additive renormalization to the quark mass (i.e., 
the residual mass) and the deviations of some amplitudes from their values 
at $N_s=\infty$, including a chirally-forbidden mixing. 
Since the cost of domain-wall simulations grows approximately linearly with 
$N_s$, it is of some importance to understand how small $N_s$ can be kept 
without occurring in large values of the residual mass. A thorough exploration 
of large regions in the two-dimensional space spanned by $N_s$ and the 
domain-wall height $M$ would be quite expensive when using Monte Carlo 
simulations, and perturbation theory remains then often the more practical 
way for gathering hints of what is happening in this space. The study 
of the dependence on $M$ and $N_s$ of various indicators of chiral violations 
is thus one of the main aims of the work.

Significant perturbative insights at finite $N_s$ beyond tree level
have been provided some years ago by Kikukawa, Neuberger and Yamada
\cite{Kikukawa:1997tf}, who diagonalized the mass matrix in the truncated 
overlap and derived 1-loop equations under certain assumptions on the gauge 
fields. We improve here on this by providing numerical results for the residual
mass. What we present is a complete 1-loop calculation of such radiative 
effects, with the aim of seeing how $m_{res}$, and other quantities which 
can act as indicators of chiral symmetry breaking, behave when $N_s$
and $M$ change. 

Other methods have also been used to gain insights on these effects.
Recently Christ \cite{Christ:2005xh} has investigated the residual mass 
analytically by looking at the eigenfunctions of the five-dimensional transfer 
matrix \cite{Furman:1994ky}. Building on the understanding of the localization 
properties of the domain-wall modes, characterized by their mobility edge
$\lambda_c$ \cite{Golterman:2003qe,Golterman:2004cy,Golterman:2005fe} 
(for recent investigations see 
\cite{Svetitsky:2005qa,Antonio:2005wm,Draper:2005mh}),
the leading effects were estimated as 
\begin{equation}
m_{res} \sim R^4_e \, \rho_e (\lambda_c) \, \frac{\exp{(-\lambda_c N_s)}}{N_s}
+ R^4_l \, \rho_l (0) \, \frac{1}{N_s} ,
\label{eq:transmat}
\end{equation}
where $\rho$ is the density per unit spacetime volume of the eigenvalues 
of the logarithm of the transfer matrix, and $l$ and $e$ stand for localized
and extended modes with average size $R$ respectively. The energy threshold
from localized to extended modes is given by the mobility edge $\lambda_c$, 
which is then also responsible for the speed with which the chiral violations 
decay with $N_s$.

This article is organized as follows. In Sect. \ref{sec:pt} we introduce all
propagator functions which are necessary for the perturbative calculations 
at finite $N_s$, and in Sect. \ref{sec:rf} we analyze the one-loop 
renormalization of quark amplitudes in this theory. Since in this article 
we restrict ourselves to the calculation of finite diagrams, the treatment 
of divergences in the theory at finite $N_s$ is left for a future work. 
We have however computed the coefficients of the divergent terms in a few 
cases, and we have seen that they are in general not equal to their continuum 
values. In fact, they depend on $N_s$ and on the height of the domain wall, 
$M$, as we show in Sect. \ref{sec:divergences}, where we also briefly discuss 
the implications of this finding. In Sect. \ref{sec:rm} we then present the 
computation of the residual mass and encounter another feature of calculations 
at finite $N_s$, namely that gauge invariance is lost. In Sect. \ref{sec:va} 
we show the results at finite $N_s$ for the difference between the vector and 
axial-vector renormalization constants, which should be zero at infinite $N_s$.
In Sect. \ref{sec:d1} we finally present a power-divergent mixing due to the 
breaking of chiral symmetry, for an operator which describes polarized 
parton distributions, and in Sect. \ref{sec:borders} we discuss what happens 
near the borders of allowed values of $M$, $M\to 0$ and $M\to 2$, before making
in Sect. \ref{sec:concl} some concluding remarks.

\section{Perturbation theory}
\label{sec:pt}

We work with the standard formulation of domain-wall fermions of Shamir 
\cite{Shamir:1993zy}, 
\begin{eqnarray}
S^{DW}_q &=& \sum_x \sum_{s=1}^{N_s} \Bigg[ 
\frac{1}{2} \sum_\mu \Big( \overline{\psi}_s(x)
(\gamma_\mu - r) U_\mu (x) \psi_s(x+\hat{\mu}) 
\nonumber \\
&& - \overline{\psi}_s(x)
(\gamma_\mu + r) U^\dagger_\mu (x-\hat{\mu}) \psi_s(x-\hat{\mu}) \Big) 
\nonumber \\
&& + \Big( \overline{\psi}_s(x) P_+ \psi_{s+1}(x)
         + \overline{\psi}_s(x) P_- \psi_{s-1}(x) \Big)
\nonumber \\
&& + (M -1 +4r) \overline{\psi}_s(x) \psi_s(x) \Bigg] \\
&& + m \sum_x \Big( \overline{\psi}_{N_s}(x) P_+ \psi_1(x) 
+ \overline{\psi}_1(x) P_- \psi_{N_s}(x) \Big) , \nonumber 
\label{eq:dwaction}
\end{eqnarray}
where we put $r=-1$, that is the Wilson term is added to the action with minus 
the conventional sign. The height of the domain wall, or Dirac mass, $M$, 
at tree level satisfies $0<M<2$, so that the correct pattern of chiral modes 
(with no doublers) is attained when $N_s\to\infty$: in fact for $M<0$ there is 
no chiral mode, while for $M>2$ there are four of them (and even more when $M$ 
is further increased). The chiral projectors are $P_\pm = (1 \pm \gamma_5)/2$.
Here and in most of the paper we put $a=1$, but in some contexts, like when 
discussing the residual mass, the lattice spacing will be explicitly shown. 

We refer to \cite{Aoki:1998vv,Blum:1999xi,Capitani:2005vb} for the Feynman 
rules which derive from this domain-wall action in the limit $N_s\to\infty$. 
In this article we always work at finite $N_s$, and the expressions that we 
have to use for the quark propagators are then different. They were partially 
derived in \cite{Aoki:1997xg}, and here we compute the remaining functions 
and provide the complete set of required propagators.

In (four-dimensional) momentum space the domain-wall Dirac operator
has the form 
\begin{eqnarray}
D_{st} (p) &=& \delta_{s,t} \, \sum_\mu i \gamma_\mu \sin p_\mu \\
&+& (W^+_{st} (p) + mM^+_{st}) P_+ + (W^-_{st} (p) +mM^-_{st}) P_- , \nonumber
\end{eqnarray}
where the mass matrix is given by 
\begin{eqnarray}
W^\pm_{st}(p) &=& -W(p) \, \delta_{s,t} + \delta_{s\pm 1,t} , \\
M^+_{st} &=& \delta_{s,N_s} \, \delta_{t,1} , \\
M^-_{st} &=& \delta_{s,1} \, \delta_{t,N_s} , 
\end{eqnarray}
and
\begin{equation}
W(p) = 1 - M - 2 r \sum_\lambda \sin^2 \frac{p_\lambda}{2} .
\end{equation}
In more explicit form, we have
\begin{eqnarray}
W^+(p) &=& \pmatrix{
-W(p) & 1      &        &       \cr
      & -W(p)  & \ddots &       \cr
      &        & \ddots & 1     \cr
      &        &        & -W(p) \cr} ,
\label{eq:massmatrix1} \\
W^-(p) &=& \pmatrix{
-W(p) &        &        &       \cr
 1    & -W(p)  &        &       \cr
      & \ddots & \ddots &       \cr
      &        & 1      & -W(p) \cr} ,
\label{eq:massmatrix2} \\
M^+ &=& \pmatrix{
  &  &   \cr
  &  &   \cr
1 &  &   \cr} ,
\label{eq:massmatrix3} \\
M^- &=& \pmatrix{
  &  & 1 \cr
  &  &   \cr
  &  &   \cr} .
\label{eq:massmatrix4} 
\end{eqnarray}
In practical terms this theory looks like having several flavors of lattice 
Dirac fermions, which are mixed in a very special way so that a large mass 
hierarchy is generated. At the end this theory indeed contains one chiral mode 
which is nearly massless together with $N_s -1$ heavy fermions.

In this work we only consider the case in which no explicit mass term
appears in the Lagrangian ($m=0$). The tree-level 5-dimensional
quark propagator is then given by
\begin{widetext}
\begin{equation}
\langle \psi_s (-p) \overline\psi_t (p)\rangle
 = \sum_u \Big[ \Big( - i \gamma_\mu \sin p_\mu \,\delta_{s,u} 
                      +W^-_{su}(p)\,\Big)\, G^R_{ut}(p) \,P_+ 
 + \Big( - i \gamma_\mu \sin p_\mu \,\delta_{s,u} 
                      +W^+_{su}(p)\,\Big)\, G^L_{ut}(p) \,P_- \Big] ,
\label{eq:psipsi}
\end{equation}
where the expressions of the functions $G^R(p)$ and $G^L(p)$ are 
\cite{Aoki:1997xg}
\begin{eqnarray}
G^R_{st}(p) &=& \frac{A(p)}{F(p)} \Bigg[ 
(1-W(p)e^{-\alpha(p)}) (e^{-2N_s\alpha(p)}-1) e^{(s+t)\alpha (p)}
+ 2 W(p) \sinh (\alpha (p)) (e^{(s-t)\alpha (p)}+e^{-(s-t)\alpha (p)}) 
\nonumber \\
&&+ (1-W(p)e^{\alpha(p)}) (1-e^{2N_s\alpha(p)}) e^{-(s+t)\alpha (p)} 
\Bigg] 
+ A(p) \Big( e^{(N_s-|s-t|)\alpha(p)} + e^{-(N_s-|s-t|)\alpha(p)} \Big) ,
\end{eqnarray}
\begin{eqnarray}
G^L_{st}(p) &=& \frac{A(p)}{F(p)} \Bigg[ 
(e^{-2\alpha(p)}-W(p)e^{-\alpha(p)}) (e^{-2N_s\alpha(p)}-1) e^{(s+t)\alpha (p)}
+ 2 W(p) \sinh (\alpha (p)) (e^{(s-t)\alpha (p)}+e^{-(s-t)\alpha (p)}) 
\nonumber \\
&&
+(e^{2\alpha(p)}-W(p)e^{\alpha(p)}) (1-e^{2N_s\alpha(p)}) e^{-(s+t)\alpha (p)}
\Bigg] 
+ A(p) \Big( e^{(N_s-|s-t|)\alpha(p)} + e^{-(N_s-|s-t|)\alpha(p)} \Big) .
\end{eqnarray}
\end{widetext}
In these expressions the quantity $\alpha(p)$ appears, which is defined by the 
positive solution of the equation \cite{Narayanan:1992wx,Shamir:1993zy}
\begin{equation}
\cosh (\alpha(p)) = \frac{1+W^2(p)+\sum_\lambda \sin^2 p_\lambda}{2|W(p)|} ,
\end{equation}
and one uses the abbreviations
\begin{eqnarray}
A(p) &=& \frac{1}{2W(p) \, \sinh (\alpha (p))} \, 
        \frac{1}{2 \sinh (N_s \alpha (p))} , \\
F(p) &=& e^{N_s \alpha(p)} (1-W(p)e^{\alpha(p)}) \\
     & & - e^{-N_s \alpha(p)} (1-W(p)e^{-\alpha(p)}) \nonumber .
\end{eqnarray}
When $W$ is negative, a situation which arises only when $1 < M < 2$
if the momentum is small enough, the propagator is given by the above 
equations with the replacements 
\begin{eqnarray}
W &\rightarrow& - |W|, \\
e^{\pm \alpha} &\rightarrow& -e^{\pm \alpha} ,
\end{eqnarray}
which imply that also $\sinh \alpha$ changes sign.

The standard ``physical'' quark fields used both in Monte Carlo simulations 
and in perturbative calculations are given by
\begin{eqnarray}
q(x) &=& P_+ \psi_1 (x) + P_- \psi_{N_s} (x) \\
\overline q(x) &=& \overline\psi_{N_s} (x) P_+ + \overline\psi_1 (x) P_- .
\end{eqnarray}
Strictly speaking these fields do not correspond exactly to the chiral modes, 
which should be instead eigenvectors of the mass matrix, like
\begin{equation}
\chi_0 (x) = \sqrt{1-w_0^2} \sum_s (P_+ w_0^{s-1} \psi_s (x) 
                            + P_- w_0^{N_s-s} \psi_s (x)) ,
\label{eq:chi0}
\end{equation}
where one calls
\begin{equation}
w_0 = W(0) = 1 - M .
\label{eq:w0}
\end{equation}
The physical quark field $q(x)$ is however more convenient to use than 
$\chi_0 (x)$, given that $w_0$ undergoes already at one loop a renormalization
deriving from the additive correction to the domain-wall height $M$ 
(see Eq. (\ref{eq:addrenormw0}) below). Thus one instead takes $q(x)$ to 
represent the physical zero modes of the theory. Moreover, at finite $N_s$ an 
additional issue about $\chi_0(x)$ would arise, because this field is an 
eigenvector of the mass matrix only up to terms of order 
$N_s \, e^{-N_s\alpha(0)}$ \cite{Aoki:1997xg,Kikukawa:1997tf,Blum:1999xi}.

The computation of matrix elements involving physical states and operators 
requires the introduction of additional propagators which connect the 
4-dimensional physical quark fields with the 5-dimensional quark fields which
appear in the Lagrangian. We have here derived the expressions of these 
propagators for the case of finite $N_s$, and they are given by
\begin{widetext}
\begin{eqnarray}
\langle q (-p) \overline\psi_s (p)\rangle &=& 
  P_+ \langle \psi_1     (-p) \overline\psi_s (p)\rangle 
+ P_- \langle \psi_{N_s} (-p) \overline\psi_s (p)\rangle \nonumber \\
&=& \Big( \frac{i \gamma_\mu \sin p_\mu}{E(p)} 
  + e^{-N_s\alpha(p)} \, \frac{2 W(p) \sinh (\alpha (p))}{
       E(p) \, \big(1-e^{-2N_s\alpha(p)} \big)} \Big) \nonumber \\
 && \qquad \times \Big( 
  \big(e^{-(N_s-s)\alpha(p)} -e^{-2N_s\alpha(p)} e^{(N_s-s)\alpha(p)}\big) P_+ 
 +\big(e^{-(s-1)\alpha(p)} -e^{-2N_s\alpha(p)} e^{(s-1)\alpha(p)}\big) P_-\Big)
\nonumber \\
 && - \frac{1}{1-e^{-2N_s\alpha(p)}} \, e^{-\alpha(p)}  \, \Big( 
   \big( e^{-(s-1)\alpha(p)} - e^{-2(N_s-1)\alpha(p)} 
   e^{(s-1)\alpha(p)} \big) P_+ \label{eq:qpsi} \\
 && \qquad \qquad \qquad \qquad \qquad + \big( e^{-(N_s-s)\alpha(p)}     
  - e^{-2(N_s-1)\alpha(p)} e^{(N_s-s)\alpha(p)} \big) P_- \Big) 
  \nonumber , \\
\langle \psi_s (-p) \overline q (p)\rangle &=& 
  \langle \psi_s (-p) \overline\psi_1     (p)\rangle P_-
+ \langle \psi_s (-p) \overline\psi_{N_s} (p)\rangle P_+ \nonumber \\
&=& \Big( 
  \big(e^{-(N_s-s)\alpha(p)} -e^{-2N_s\alpha(p)} e^{(N_s-s)\alpha(p)}\big) P_- 
 +\big(e^{-(s-1)\alpha(p)} -e^{-2N_s\alpha(p)} e^{(s-1)\alpha(p)}\big) P_+\Big)
 \nonumber \\
 && \qquad \times 
\Big( \frac{i \gamma_\mu \sin p_\mu}{E(p)} 
  + e^{-N_s\alpha(p)} \, \frac{2 W(p) \sinh (\alpha (p))}{
       E(p) \, \big(1-e^{-2N_s\alpha(p)} \big)} \Big) 
\nonumber \\
 && - \frac{1}{1-e^{-2N_s\alpha(p)}} \, e^{-\alpha(p)} \, \Big( 
   \big( e^{-(s-1)\alpha(p)} - e^{-2(N_s-1)\alpha(p)}  
   e^{(s-1)\alpha(p)} \big) P_- \label{eq:psiq}\\
 && \qquad \qquad \qquad \qquad \qquad + \big( e^{-(N_s-s)\alpha(p)}   
  - e^{-2(N_s-1)\alpha(p)} e^{(N_s-s)\alpha(p)} \big) P_+ \Big) 
  \nonumber ,
\end{eqnarray}
where we have defined
\begin{equation}
E(p) = 1 - W(p)e^{\alpha(p)} 
  - e^{-2N_s \alpha(p)} \big(1-W(p)e^{-\alpha(p)}\big) .
\end{equation}
We have also numerically checked the above expressions and the correctness of  
their implementation in our computer codes by verifying the validity 
for each $s$ (for various choices of $N_s$) of the identities
\begin{eqnarray}
G^L_{1s}(p)  =  G^L_{s1}(p)  &=&  -\frac{1}{E(p)} \, 
\big(e^{-(s-1)\alpha(p)} -e^{-2N_s\alpha(p)} e^{(s-1)\alpha(p)}\big) , \\
G^R_{N_ss}(p)  =  G^R_{sN_s}(p)  &=&  -\frac{1}{E(p)} \,
\big(e^{-(N_s-s)\alpha(p)} -e^{-2N_s\alpha(p)} e^{(N_s-s)\alpha(p)}\big) , \\
\sum_t W^+_{N_st}(p) G^L_{ts}(p)  =  \sum_t W^-_{st}(p) G^R_{tN_s}(p)  &=& 
- \frac{1}{1-e^{-2N_s\alpha(p)}} \, e^{-\alpha(p)} \, 
\big(e^{-(N_s-s)\alpha(p)} -e^{-2(N_s-1)\alpha(p)} e^{(N_s-s)\alpha(p)}\big) \\
&& + e^{-N_s\alpha(p)} \, \frac{2 W(p) \sinh (\alpha (p))}{
       E(p) \, \big(1-e^{-2N_s\alpha(p)} \big)} \,
\big(e^{-(s-1)\alpha(p)} -e^{-2N_s\alpha(p)} e^{(s-1)\alpha(p)}\big) , 
\nonumber \\
\sum_t W^-_{1t}(p) G^R_{ts}(p) =  \sum_t W^+_{st}(p) G^L_{t1}(p)  &=& 
- \frac{1}{1-e^{-2N_s\alpha(p)}} \, e^{-\alpha(p)} \, 
\big(e^{-(s-1)\alpha(p)} -e^{-2(N_s-1)\alpha(p)} e^{(s-1)\alpha(p)}\big) \\ 
&& + e^{-N_s\alpha(p)} \, \frac{2 W(p) \sinh (\alpha (p))}{
       E(p) \, \big(1-e^{-2N_s\alpha(p)} \big)} \,
\big(e^{-(N_s-s)\alpha(p)} -e^{-2N_s\alpha(p)} e^{(N_s-s)\alpha(p)}\big) , 
\nonumber 
\end{eqnarray}
which relate the propagators $\langle q (-p) \overline\psi_s (p)\rangle$ and 
$\langle \psi_s (-p) \overline q (p)\rangle$ to
$\langle \psi_s (-p) \overline\psi_t (p)\rangle$.

The calculation of perturbative amplitudes also requires the knowledge of
the expressions of these new propagators for small momentum. In this limit 
we obtain 
\begin{eqnarray}
\langle q (-p) \overline \psi_s (p)\rangle_c &=& 
  -\frac{1-w_0^2}{1 - w_0^{2N_s}} \, 
 \frac{i\slash{p} + w_0^{N_s} (1-w_0^2)}{p^2 + w_0^{2N_s}(1-w_0^2)^2} \, 
 \Big( \big(w_0^{N_s-s} - w_0^{2N_s} w_0^{-(N_s-s)}\big) P_+ 
     + \big(w_0^{s-1} - w_0^{2N_s} w_0^{-(s-1)}\big) P_- \Big)
 \nonumber \\
&& - \frac{1}{1-w_0^{2N_s}} \, w_0 \, 
\Big( \big(w_0^{s-1} - w_0^{2(N_s-1)} w_0^{-(s-1)} \big) P_+
    + \big(w_0^{N_s-s} - w_0^{2(N_s-1)} w_0^{-(N_s-s)} \big) P_- 
 \Big) , \label{eq:qpsi0}\\
\langle \psi_s (-p) \overline q (p)\rangle_c &=& 
  -\frac{1-w_0^2}{1 - w_0^{2N_s}} \, 
 \Big( \big(w_0^{N_s-s} - w_0^{2N_s} w_0^{-(N_s-s)}\big) P_- 
     + \big(w_0^{s-1} - w_0^{2N_s} w_0^{-(s-1)}\big) P_+ \Big) \,
 \frac{i\slash{p} + w_0^{N_s} (1-w_0^2)}{p^2 + w_0^{2N_s}(1-w_0^2)^2}  
 \nonumber \\
&& - \frac{1}{1-w_0^{2N_s}} \, w_0 \, 
\Big( \big(w_0^{s-1} - w_0^{2(N_s-1)} w_0^{-(s-1)} \big) P_-
    + \big(w_0^{N_s-s} - w_0^{2(N_s-1)} w_0^{-(N_s-s)} \big) P_+ 
 \Big) , \label{eq:psiq0} 
\end{eqnarray}
\end{widetext}
where $w_0$ is defined in Eq. (\ref{eq:w0}). 
Since $w_0 = e^{-\alpha(0)}$, it is easy to see that the terms which are 
proportional to $w_0^{N_s} = e^{-N_s\alpha(0)}$ rapidly approach zero when
$N_s$ becomes large. In the derivation of the above formulae we have used 
the useful small momentum expansions
\begin{eqnarray}
1 - W(p)e^{\alpha(p)} &=& - \frac{p^2}{1 -w_0^2} , \\
1 - W(p)e^{-\alpha(p)} &=& 1 -w_0^2 
              - w_0 \, \frac{1 -w_0 -w_0^2}{1 -w_0^2} \, p^2 
\end{eqnarray}
and
\begin{equation}
e^{-N_s \alpha(p)} \, \frac{2 W(p) \sinh (\alpha (p))}{E(p)} = 
- \frac{ w_0^{N_s} (1 -w_0^2)^2 }{p^2 + w_0^{2N_s} (1 -w_0^2)^2} ,
\end{equation}
and we have dropped all terms of order $p^2 w_0^{N_s}$ and higher, which are
much smaller than either of the factors $p^2$ or $w_0^{N_s}$ alone, and are 
not relevant when $p\to 0$. It is easy to check that all the propagators 
that we have derived in this Section reduce to the expressions used in the 
calculations of Refs. \cite{Aoki:1998vv,Blum:1999xi,Capitani:2005vb} when $N_s$
is large. Notice also that the function $E(p)$ introduced here tends in this 
approximation to the function $F(p)$ as defined in those articles. The above 
expressions for the small momentum propagators can also be used for $M>1$ 
without further modifications (as we have also numerically checked).

Finally, we also need the function that describes the propagation of the
physical fields alone. This is given by 
\begin{eqnarray}
\langle q (-p) \overline q (p)\rangle &=& \frac{1}{E(p)} \, \Big(
i \gamma_\mu \sin p_\mu \, (1-e^{-2N_s\alpha(p)}) \label{eq:qq} \\
&& \quad + e^{-N_s\alpha(p)} \cdot 2 W(p) \sinh (\alpha (p)) \Big) , \nonumber
\end{eqnarray}
which in the limit of small momentum becomes
\begin{equation}
\langle q (-p) \overline q (p)\rangle_c = - (1-w_0^2) \, 
\frac{i\slash{p} + w_0^{N_s} (1-w_0^2)}{p^2 + w_0^{2N_s}(1-w_0^2)^2} .
\label{eq:qq0}
\end{equation}
It is interesting to see that it is also possible to calculate 
$\langle q (-p) \overline q (p)\rangle$ in an alternative way directly from 
the 5-dimensional propagator of Eq.~(\ref{eq:psipsi}):
\begin{eqnarray}
\langle q \overline q \rangle &=& 
     P_+ \langle \psi_1 \overline\psi_{N_s} \rangle P_+ 
   + P_+ \langle \psi_1 \overline\psi_1 \rangle P_- \nonumber \\
&& + P_- \langle \psi_{N_s} \overline\psi_{N_s} \rangle P_+ 
   + P_- \langle \psi_{N_s} \overline\psi_1 \rangle P_- \nonumber \\
&=& -\frac{1}{2} \, \Big[ i \gamma_\mu \sin p_\mu 
\Big( G^R_{N_sN_s} \, P_+ + G^L_{11} \, P_- \Big) \nonumber \\
&& + W \, \Big( G^R_{1N_s} \, P_+ + G^L_{N_s1} \, P_- \Big) \Big] .
\end{eqnarray}
The fact that in this way we obtain again the result of Eq.~(\ref{eq:qq0}) 
provides a good check of the above formulae.

Domain-wall fermions present thus at finite $N_s$ some new peculiar features. 
Although the theory which we have started from is described by a Lagrangian 
of massless quarks, the propagator of the physical quark field, 
$\langle q (-p) \overline q (p)\rangle_c$, acquires when $N_s$ is kept finite 
a nonvanishing mass term, which at tree level is given by
\begin{equation}
a \, m_{res}^{(0)} = - w_0^{N_s} (1-w_0^2) = - (1-M)^{N_s} \, M(2-M) ,
\label{eq:mres_tree}
\end{equation}
as can be seen from the general expression of a fermion propagator of mass 
$\mu$ for small momentum in Euclidean space:
\begin{equation}
\frac{-i\slash{p} + \mu}{p^2 + \mu^2} = \frac{1}{i\slash{p} + \mu} .
\end{equation}
We readily see that this tree-level residual mass vanishes when $N_s=\infty$. 
We will only consider even values of $N_s$, in which case the fermion 
determinant can be proven to be positive (so that the square root of the 
two-flavor theory is well defined and an odd number of dynamical flavors 
can be simulated). Then $m_{res}^{(0)}$ is always a negative quantity. 
With our calculation we have thus reproduced, up to a sign, the result for 
$m_{res}^{(0)}$ found in 
\cite{Shamir:1993zy,Vranas:1997da,Vranas:1997ib,Kikukawa:1997tf,Blum:1999xi}.
This was derived by considering the quadratic operator $D^\dagger D$,
which could perhaps explain the sign discrepancy.

We will see that when the one-loop corrections are taken into account, 
the residual mass changes sign and becomes positive.

If we had used the chiral mode $\chi_0$, Eq.~(\ref{eq:chi0}), the propagator 
for small momentum would have been given by
\begin{eqnarray}
\langle (1-w_0^2) \, \chi_0 \overline \chi_0 \rangle_c &=& 
- \frac{1-w_0^2}{p^2 + w_0^{2N_s}(1-w_0^2)^2} \nonumber \\
&& \times \Bigg[ i\slash{p} + w_0^{N_s} (1-w_0^2) (1 - 2w_0^{2N_s})\nonumber \\
&& \qquad - N_s \frac{w_0^{2N_s} (1-w_0^2)^2}{1 - w_0^{2N_s}} \Bigg] , 
\label{eq:propchi0}
\end{eqnarray}
which compared to $\langle q \overline q \rangle_c$ has correction terms 
of higher order in $w_0^{N_s}$ in the numerator.
We remind however that $\chi_0$ deviates from what would be the exact 
chiral mode for finite $N_s$ by terms of order $N_s \, w_0^{N_s}$. 
The real physical propagator would then also present further corrections terms,
and it could be that the expression in Eq. (\ref{eq:propchi0}) would get 
simpler.

To complete the setup of our calculations, we also recall that we use the 
plaquette action in a general covariant gauge, where the gluon propagator 
is given by
\begin{equation}
G_{\mu\nu}(k) = \frac{1}{4\sum_\rho \sin^2 \frac{k_\rho}{2}}
\Bigg( \delta_{\mu\nu} - (1-\alpha) \frac{4 \sin \frac{k_\mu}{2} 
\sin \frac{k_\nu}{2}}{4\sum_\lambda \sin^2 \frac{k_\lambda}{2}} \Bigg) ,
\end{equation}
so that $\alpha=1$ and $\alpha=0$ correspond to the Feynman and Landau gauges 
respectively. The QCD vertices that we need are the usual ones, and 
(apart from color factors) they have the form 
\begin{eqnarray}
V^{(1)}_\mu (p) &=& - g_0 \, \Big( i \gamma_\mu \cos \frac{p_\mu}{2}
              - \sin \frac{p_\mu}{2} \Big)  \\
V^{(2)}_{\mu\nu} (p) &=& \frac{1}{2} \, g_0^2 \, \Big( i \gamma_\mu 
   \sin \frac{p_\mu}{2} + \cos \frac{p_\mu}{2} \Big) \cdot \delta_{\mu\nu}
\end{eqnarray}
for the interaction of the quark current with one gluon and two gluons 
respectively, where $p$ in this case stands for the sum of the incoming 
and outgoing quark momenta.

\section{Renormalization at finite $N_s$}
\label{sec:rf}

Let us now investigate the properties of the renormalization of the self-energy
of a massless quark at finite $N_s$. We first notice that the one-loop 
propagator of the physical field can be written, according to the general form 
of the external legs $\langle q (-p) \overline\psi_s (p)\rangle_c$ and 
$\langle \psi_s (-p) \overline q (p)\rangle_c$, as 
\begin{widetext}
\begin{eqnarray}
\langle q (-p) \overline q (p)\rangle_{1~loop} &=&
\frac{1-w_0^2}{i\slash{p}-w_0^{N_s}(1-w_0^2)} +
\frac{1-w_0^2}{i\slash{p}-w_0^{N_s}(1-w_0^2)} \, 
\Sigma_q (p) \, \frac{1-w_0^2}{i\slash{p}-w_0^{N_s}(1-w_0^2)}  
\nonumber \\
&=& \frac{1-w_0^2}{i\slash{p}-w_0^{N_s}(1-w_0^2)
- (1-w_0^2) \, \Sigma_q (p) } ,
\end{eqnarray}
where 
\begin{eqnarray}
\Sigma_q (p) = & \sum_{s,t} & \frac{1}{1-w_0^{2N_s}} \, \Big[ 
   \big(w_0^{N_s-s} - w_0^{2N_s} w_0^{-(N_s-s)}\big) P_+ 
 + \big(w_0^{s-1} - w_0^{2N_s} w_0^{-(s-1)}\big)  P_- \nonumber \\
&& \quad -w_0 \, \frac{i\slash{p}-w_0^{N_s}(1-w_0^2)}{1-w_0^2} 
\, \Big( \big(w_0^{s-1} - w_0^{2(N_s-1)} w_0^{-(s-1)} \big) P_+ 
       + \big(w_0^{N_s-s} - w_0^{2(N_s-1)} w_0^{-(N_s-s)} \big) P_- 
 \Big) \Big] \nonumber \\
&& \cdot \Sigma_{st} (p) \label{eq:sigmaq} \\ 
&& \cdot \frac{1}{1-w_0^{2N_s}} \, \Big[ 
   \big(w_0^{N_s-t} - w_0^{2N_s} w_0^{-(N_s-t)}\big) P_- 
 + \big(w_0^{t-1} - w_0^{2N_s} w_0^{-(t-1)}\big)  P_+ \nonumber \\
&& \quad - w_0 \, \Big( \big(w_0^{t-1} - w_0^{2(N_s-1)} w_0^{-(t-1)} \big) P_- 
       + \big(w_0^{N_s-t} - w_0^{2(N_s-1)} w_0^{-(N_s-t)} \big) P_+ 
 \Big) \, \frac{i\slash{p}-w_0^{N_s}(1-w_0^2)}{1-w_0^2}
\Big] \nonumber 
\end{eqnarray}
contributes to the $g_0^2$ order and the functions $G^R_{st}$ and $G^L_{st}$
give contributions to $\Sigma_{st} (p)$ only. The general form of 
$\Sigma_q (p)$ is 
\begin{equation}
\Sigma_q (p) =  \frac{\bar g^2}{1-w_0^2} \, \Big[ \frac{\Sigma_0}{a}  
+ i\slash{p} \, \Big( c_{\Sigma_1}^{(N_s,M)} \log a^2 p^2 + \Sigma_1 \Big)
- \big( i\slash{p}-w_0^{N_s}(1-w_0^2) \big) \, 
\frac{2w_0}{1-w_0^2} \, \Sigma_3 \Big] , 
\label{eq:seint}
\end{equation}
\end{widetext}
where from now on we call for brevity $\bar g^2 = (g_0^2/16 \pi^2)\, C_F$ 
(with $C_F=(N_c^2-1)/2N_c$ for the $SU(N_c)$ gauge group). We can observe that
$\Sigma_q $ differs in a few aspects from the corresponding expression for 
infinite $N_s$. In fact, apart from the different coefficient of the 
logarithmic term, which now depends on $N_s$ and $M$ (see Section 
\ref{sec:divergences}), and the slightly different coefficient of $\Sigma_3$, 
it also contains a totally new contribution proportional to $1/a$, called 
$\Sigma_0$, which is a mass correction term. The one-loop radiatively 
induced mass is indeed given by
\begin{equation}
a \, m_{res}^{(1)} = - w_0^{N_s}(1-w_0^2) - \bar g^2 \, \Sigma_0 ,
\label{eq:mres}
\end{equation}
as can be easily seen when the one-loop correction to the quark 
propagator is cast in the same form as its tree-level expression:
\begin{eqnarray}
\langle q (-p) \overline q (p)\rangle_{1~loop}
&=& \frac{1-w_0^2}{i\slash{p} -w_0^{N_s}(1-w_0^2) -(1-w_0^2) \, \Sigma_q (p)}
\nonumber \\ 
&=& \frac{1-w_0^2}{i\slash{p} \, Z_2^{-1} + m_{res}^{(1)}} \, Z_w \, . 
\label{eq:o-loop}
\end{eqnarray}
This $O(a^{-1})$ critical mass, reminiscent of the analogous quantity for 
Wilson fermions, vanishes when the theory describes exact chiral fermions, 
that is at infinite $N_s$, but is different from zero when computations 
are done at any finite $N_s$. This means that in the latter case the 
$\Sigma_0$ contribution generates a finite additive renormalization to the 
quark mass, which can represent a measure of chirality-breaking effects.
We associate this perturbative critical mass $m_{res}$ with the residual mass 
which in Monte Carlo simulations is computed by looking at the explicit chiral 
symmetry breaking term in the axial Ward identities.

The results above should not come as a surprise, since after all we are working
here with a theory of $N_s$ Wilson fermions, $N_s-1$ of which are heavy states 
(not counting the doublers), and the necessity of a fine tuning to some 
critical mass comes out naturally. As in the case of Wilson fermions, where 
the hopping parameter is renormalized away from its tree level value 1/8, 
in order to obtain a massless pion in domain-wall simulations the quark mass 
must be tuned to a nonzero number, which at one loop is given by the critical 
mass given above. This defines the chiral limit when no explicit mass term
appears in the Lagrangian. Of course higher loops and nonperturbative effects 
give further contributions to the shift of the critical mass. From a practical 
point of view, it is interesting to see how small this critical mass is, 
together with its dependence on $N_s$ (and $M$). 
In the free case the numerical results for $m_{res}$ according 
to Eq.~(\ref{eq:mres_tree}) are collected in Table \ref{tab:residual_tree}
(where they have already been multiplied for $16 \pi^2$, so as to ease the 
comparisons with the one-loop results for the critical mass presented 
in the next Section). We expect that the critical mass substantially decreases 
when $N_s$ becomes large, as can be verified at one loop from Tables 
\ref{tab:residual1} to \ref{tab:residual2l} in the next Section. 

The other quantities appearing in the last line of Eq.~(\ref{eq:o-loop}) are
\begin{equation}
Z_2 =  1 + \bar g^2 \, 
\Big( c_{\Sigma_1}^{(N_s,M)} \log a^2 p^2 + \Sigma_1 \Big) ,
\end{equation}
which is the quark wave function renormalization factor, and
\begin{equation}
Z_w =  1 - \frac{2w_0}{1-w_0^2} \, \bar g^2 \, \Sigma_3
=  1 + \bar g^2 \, z_w , 
\end{equation}
which represents an additive renormalization to $w_0$ and hence to the 
domain-wall height $M$ \cite{Aoki:1998vv}, as can be inferred from 
\begin{equation}
(1-w_0^2) \, Z_w = 1 - \big( w_0 + \bar g^2 \, \Sigma_3 \big)^2 + O(\bar g^4).
\label{eq:addrenormw0}
\end{equation}
There is indeed no chiral symmetry which can protect this mass, even at 
$N_s=\infty$. The additive renormalization to $M$ can be traced back, in the 
damping factors $\langle q (-p) \overline \psi_s (p)\rangle_c$ and
$\langle \psi_s (-p) \overline q (p)\rangle_c$ in Eq.~(\ref{eq:sigmaq}), 
to the terms which are proportional to $i\slash{p}$. 
Notice how these terms are also proportional to $w_0^{N_s}(1-w_0^2)$, 
which although being of a different order in $a$ is required for the correct 
recasting of the one-loop propagators in the form of Eq.~(\ref{eq:o-loop}).

The renormalization of a composite operator $\overline q(x) \, O \, q(x)$ 
which is multiplicatively renormalizable can also be expressed in a simple way.
Again, by looking at the general form of the propagators the one-loop matrix 
element of such an operator between ``physical'' quark states can be written as
\begin{eqnarray}
\langle \, ( \, \overline q O q \, ) \, q \overline q \, \rangle_{1~loop} &=& 
\frac{1-w_0^2}{i\slash{p}-w_0^{N_s}(1-w_0^2)} \cdot A_O (p) \cdot O 
\nonumber \\
&& \times \frac{1-w_0^2}{i\slash{p}-w_0^{N_s}(1-w_0^2)} ,
\end{eqnarray}
where $A_O (p)$ contains the contribution of the damping factors. 
For a logarithmically divergent operator it takes the form
\begin{equation}
A_O (p) = \bar g^2 \Big( - \gamma_O^{(N_s,M)} \log a^2 p^2  + B_O \Big),
\label{eq:matel1}
\end{equation}
where the anomalous dimension turns out in general to be a function of $N_s$ 
and $M$, even at lowest order.

\section{Divergences at finite $N_s$}
\label{sec:divergences}

As we have anticipated in the previous Section, in the case of the self-energy,
Eq.~(\ref{eq:seint}), and of a divergent operator, Eq.~(\ref{eq:matel1}), 
the coefficients of the logarithmic divergences turn out to depend 
on $N_s$ and $M$. It is only when $N_s=\infty$ that they become equal to the 
ones calculated in the continuum. We give here some examples of this phenomenon
by computing a few of these coefficients. 

The divergence of the half-circle diagram of the self-energy comes 
from the terms which are of first order in $p$ in the integral
\begin{widetext}
\begin{eqnarray}
 2i \, \int \, \frac{d^4 k}{(2\pi)^4} \, \frac{1-w_0^2}{(1-w_0^{2N_s})^2}  
& \sum_{\rho\lambda} \sum_{st} &
   \Big( \big(w_0^{N_s-s} - w_0^{2N_s} w_0^{-(N_s-s)}\big) P_+ 
     + \big(w_0^{s-1} - w_0^{2N_s} w_0^{-(s-1)}\big) P_- \Big) \\
&& \times \frac{k \cdot p}{(k-p)^4} \,
\Big( \delta_{\rho\lambda} -(1-\alpha)\, \frac{k_\rho k_\lambda}{(k-p)^2} \Big)
\cdot \gamma_\rho \cdot \gamma_\mu k_\mu 
  \cdot \gamma_\lambda \cdot \big( 
  \widetilde{G}^R_{st}(k) \,P_+ + \widetilde{G}^L_{st}(k) \,P_- \big) 
\nonumber \\
&& \times \Big( \big(w_0^{N_s-t} - w_0^{2N_s} w_0^{-(N_s-t)}\big) P_- 
     + \big(w_0^{t-1} - w_0^{2N_s} w_0^{-(t-1)}\big) P_+ \Big) , \nonumber 
\end{eqnarray}
where the small $k$ expansions of the functions $G^R$ and $G^L$ are given by
\begin{eqnarray}
\widetilde{G}^R_{st}(k) &=& \frac{1-w_0^2}{k^2+(m_{res}^{(0)})^2} \, 
         \left( w_0^{2N_s-s-t} - \frac{w_0^{2N_s}}{1-w_0^{2N_s}} \,  
           \left( w_0^{s-t} + w_0^{-(s-t)} \right) +O(k^2) \right) , \\
\widetilde{G}^L_{st}(k) &=& \frac{1-w_0^2}{k^2+(m_{res}^{(0)})^2} \, 
         \left( w_0^{s+t-2} - \frac{w_0^{2N_s}}{1-w_0^{2N_s}} \,  
           \left( w_0^{s-t} + w_0^{-(s-t)} \right) +O(k^2) \right) , 
\end{eqnarray}
The terms proportional to $w_0^{2N_s-s-t}$ and $w_0^{s+t-2}$, which are 
localized near the two walls and provide the leading approximation 
for large $N_s$ to the continuum coefficient, have already been given in 
\cite{Aoki:1997xg,Aoki:1998vv}.

After doing the gamma algebra and carrying out the sums in the indices $s$ and 
$t$ in the fifth dimension, we get a compact analytic expression for the 
coefficient of the logarithmic term, as a function of $N_s$ and $M$: 
\begin{equation}
c_{\Sigma_1}^{(N_s,M)} = c_{\Sigma_1}^{\infty} \cdot 
\Bigg( 1 - N_s \, w_0^{2N_s} \, \frac{1-w_0^2}{1-w_0^{2N_s}} \Bigg)
\Bigg( 1 - N_s \, w_0^{2N_s} \, \frac{1-w_0^2}{1-w_0^{2N_s}} 
+ 2 \, \frac{w_0^{2(N_s+1)}}{1-w_0^{2N_s}} \,
\Bigg( 1 - N_s \, \frac{1-w_0^2}{w_0^2 \, (1-w_0^{2N_s})} \Bigg) \Bigg) ,
\label{eq:coeffsigm1} 
\end{equation}
where $c_{\Sigma_1}^{\infty} = \alpha$ is the value of the coefficient 
in the case of exact chiral symmetry. Numerical values of 
$c_{\Sigma_1}^{(N_s,M)} $ for various choices of $N_s$ and $M$ in Feynman gauge
are shown in Table \ref{tab:div_sigma1}. 

For a generic bilinear $\overline q(x) \, \Gamma \, q(x)$, the divergence is 
obtained by computing the following integral from the vertex diagram: 
\begin{eqnarray}
- \int \, \frac{d^4 k}{(2\pi)^4} \, \frac{(1-w_0^2)^2}{(1-w_0^{2N_s})^4} 
& \sum_{\rho\lambda} \sum_{st} &
 \Big( \big(w_0^{N_s-s} - w_0^{2N_s} w_0^{-(N_s-s)}\big) P_+ 
     + \big(w_0^{s-1} - w_0^{2N_s} w_0^{-(s-1)}\big) P_- \\
&& \quad   +w_0^{N_s+1} 
  \, \Big( \big(w_0^{s-1} - w_0^{2(N_s-1)} w_0^{-(s-1)} \big) P_+ 
       + \big(w_0^{N_s-s} - w_0^{2(N_s-1)} w_0^{-(N_s-s)} \big) P_- 
 \Big) \Big) 
 \nonumber \\
&& \times \gamma_\rho \, \Big[
   \Big( \big(w_0^{N_s-s} - w_0^{2N_s} w_0^{-(N_s-s)}\big) P_- 
     + \big(w_0^{s-1} - w_0^{2N_s} w_0^{-(s-1)}\big) P_+ \Big) 
    \frac{i\slash{k}-m_{res}^{(0)}}{k^2+(m_{res}^{(0)})^2} \nonumber \\
 && \quad + \frac{w_0}{1-w_0^2}   
    \, \Big( \big(w_0^{s-1} - w_0^{2(N_s-1)} w_0^{-(s-1)} \big) P_- 
       + \big(w_0^{N_s-s} - w_0^{2(N_s-1)} w_0^{-(N_s-s)} \big) P_+  \Big) 
\Big] \nonumber \\
&& \times \, \Gamma  
 \cdot \frac{1}{(k-p)^2} 
\Big( \delta_{\rho\lambda} -(1-\alpha)\, \frac{k_\rho k_\lambda}{(k-p)^2} \Big)
\nonumber \\
&& \times \Big[ \frac{i\slash{k}-m_{res}^{(0)}}{k^2+(m_{res}^{(0)})^2} \, 
  \Big( \big(w_0^{N_s-t} - w_0^{2N_s} w_0^{-(N_s-t)}\big) P_+ 
     + \big(w_0^{t-1} - w_0^{2N_s} w_0^{-(t-1)}\big) P_- \Big) \nonumber \\
&& \quad + \frac{w_0}{1-w_0^2}   
    \, \Big( \big(w_0^{t-1} - w_0^{2(N_s-1)} w_0^{-(t-1)} \big) P_- 
       + \big(w_0^{N_s-t} - w_0^{2(N_s-1)} w_0^{-(N_s-t)} \big) P_+  \Big) 
\Big] \nonumber \\
&& \times \gamma_\lambda \, 
   \Big( \big(w_0^{N_s-t} - w_0^{2N_s} w_0^{-(N_s-t)}\big) P_- 
     + \big(w_0^{t-1} - w_0^{2N_s} w_0^{-(t-1)}\big) P_+ \nonumber \\
&& \quad   +w_0^{N_s+1} 
  \, \Big( \big(w_0^{t-1} - w_0^{2(N_s-1)} w_0^{-(t-1)} \big) P_-
       + \big(w_0^{N_s-t} - w_0^{2(N_s-1)} w_0^{-(N_s-t)} \big) P_+ 
 \Big) \Big) . \nonumber 
\end{eqnarray}
The coefficient of the logarithmic term turns out after many simplifications 
to be given by
\begin{equation}
c_{\Gamma}^{(N_s,M)} = c_{\Gamma}^{\infty} \cdot 
\Bigg( 1 - N_s \, w_0^{2N_s} \, \frac{1-w_0^2}{1-w_0^{2N_s}} \Bigg)
\Bigg( 1 - N_s \, w_0^{2N_s} \, \frac{1-w_0^2}{1-w_0^{2N_s}} 
- 4 \, w_0^{2N_s} \,
\Bigg( 1 - N_s \, \frac{1-w_0^2}{1-w_0^{2N_s}} \Bigg) \Bigg) ,
\label{eq:coeffcurrents}
\end{equation}
where $c_{\Gamma}^{\infty}$ is its continuum value, that is 
$c_S^{\infty}=c_P^{\infty}=-3+\alpha$, $c_V^{\infty}=c_A^{\infty}=-\alpha$ and 
$c_T^{\infty}=1-\alpha$. The dependence on $N_s$ and $M$ is different from the 
one of the self-energy, and thus we find that the anomalous dimensions of 
all bilinears must also depend on these parameters. Numerical values of 
$c_V^{(N_s,M)}$ are reported in Table \ref{tab:div_vertex} for the Feynman 
gauge. 

A remarkable consequence of the above formulae is that the anomalous dimension 
of the vector and axial-vector currents does not vanish anymore. The 
contributions coming from the half-circle diagram of the self-energy and from 
the vertex diagram of the vector (or axial-vector) current do not compensate 
each other, because part of the subleading terms are different. In Feynman
gauge one gets
\begin{equation}
\gamma_V^{(N_s,M)} = 2 \, w_0^{2N_s} \cdot 
\Bigg( 1 - N_s \, w_0^{2N_s} \, \frac{1-w_0^2}{1-w_0^{2N_s}} \Bigg)
\Bigg( N_s  \, \frac{1-w_0^2}{1-w_0^{2N_s}} \, 
                 \Big( 2 + \frac{1}{1-w_0^{2N_s}} \Big)
     - 2 - \frac{w_0^2}{1-w_0^{2N_s}} \Bigg),
\end{equation}
\end{widetext}
and the numerical values of this anomalous dimension are reported in Table 
\ref{tab:div_vector}. These currents have then for any finite $N_s$ an 
anomalous dimension which is different from zero. Only in the Landau gauge 
it is equal to the case of exact chiral symmetry and thus vanishes, but this 
happens just because the coefficients of the logarithms, being proportional to 
their values at $N_s=\infty$, in this gauge vanish separately for the 
self-energy and for these currents. 

The deviations of the anomalous dimension of the scalar and pseudoscalar 
densities from its $N_s=\infty$ value are given for the Feynman gauge 
in Table \ref{tab:div_scalar}.

Thus, in domain-wall fermions for any finite $N_s$ already at the one-loop 
level the anomalous dimensions of the continuum and lattice versions of an 
operator are not the same, and this raises some issues about the correct 
procedure with which one must carry out the matching of lattice operators 
to a continuum scheme in this case. In fact, the matching formula 
\cite{Martinelli:1982mw,Capitani:2002mp}
\begin{equation}
\frac{\langle q | O^{\overline{\mathrm{MS}}} | q \rangle}{
\langle q | O^{lat} | q \rangle}
= 1-\bar g^2\Big( -\gamma_O \log a^2\mu^2 +R^{lat} 
-R^{\overline{\mathrm{MS}}} \Big) 
\label{eq:1loopcontlat}
\end{equation}
is only valid provided the coefficients of the logarithmic terms in 
\begin{equation}
\langle q | O^{lat} | q \rangle = \Big( 1
+\bar g^2 \Big( -\gamma_O \log a^2p^2
+ R^{lat} \Big) \Big) \cdot
\langle q | O^{tree} | q \rangle
\label{eq:1looplat} 
\end{equation}
and
\begin{equation}
\langle q | O^{\overline{\mathrm{MS}}} | q \rangle = \Big( 1
+\bar g_{\overline{\mathrm{MS}}}^2 \Big( -\gamma_O \log \frac{p^2}{\mu^2} +
R^{\overline{\mathrm{MS}}} \Big) \Big) \cdot \langle q | O^{tree} | q \rangle 
\label{eq:1loopcont} 
\end{equation}
are the same. But this does not happen for domain-wall fermions at finite 
$N_s$, because the $\gamma_O$ in Eq.~(\ref{eq:1looplat}) must be replaced with 
a $\gamma_O^{N_s,M}$ which depends on $N_s$, and thus this logarithmic term 
cannot be combined with the one of Eq.~(\ref{eq:1loopcont}), whose coefficient 
is given by the continuum theory. In particular, the $p$ dependence cannot 
be eliminated from the matching formula.

It could well be that all this is connected to the fact that the domain-wall 
theory at finite $N_s$ is loaded with some pathologies, as it possesses
no analytical Atiyah-Singer index. It is indeed well known that in all 
formulations of chiral fermions that recover the correct quantum anomalies 
one needs (in one way or another) an infinite number of fermion fields. 
With a finite number of these fields what happens is that either the two 
chiralities cannot be completely separated or that the right anomalies cannot 
be reproduced. In domain-wall at finite $N_s$, which corresponds to a finite 
number of fermions fields, we could then anticipate something like this to 
happen. We will see in the following Section that in the theory truncated
at finite $N_s$ this is not the only strange feature which arises, but that 
gauge invariance is lost as well, even for finite quantities.

\begin{table*}[htp]
\caption{\label{tab:residual_tree}
Residual mass at tree level in lattice units (multiplied for $16 \pi^2$).}
\begin{ruledtabular}    
\begin{tabular}{|c|rrrrrrrrr|} 
$M$ & $N_s=8$ & $N_s=12$ & $N_s=16$ & $N_s=20$ & $N_s=24$ & $N_s=28$ 
    & $N_s=32$ & $N_s=48$ & $N_s=\infty$  
      \vspace{0.05cm} \\ \hline \vspace{-0.3cm} \\  
  0.1   &-12.91556 & -8.47390 & -5.55973 & -3.64774 & -2.39328 
        & -1.57023 & -1.03023 & -0.19090 &  0  \\
  0.2   & -9.53767 & -3.90663 & -1.60015 & -0.65542 & -0.26846      
        & -0.10996 & -0.04504 & -0.00127 &  0  \\
  0.3   & -4.64274 & -1.11472 & -0.26764 & -0.06426 & -0.01543 
        & -0.00370 & -0.00089 &  0.00000 &  0  \\
  0.4   & -1.69750 & -0.22000 & -0.02851 & -0.00370 & -0.00048 
        & -0.00006 & -0.00001 &  0.00000 &  0  \\
  0.5   & -0.46264 & -0.02891 & -0.00181 & -0.00011 & -0.00001 
        &  0.00000 &  0.00000 &  0.00000 &  0  \\
  0.6   & -0.08693 & -0.00223 & -0.00006 &  0.00000 &  0.00000 
        &  0.00000 &  0.00000 &  0.00000 &  0  \\
\hline    
\end{tabular}
\end{ruledtabular}
\end{table*}

\begin{table*}[htp]
\caption{\label{tab:div_sigma1}
Coefficient of the logarithmic term for $\Sigma_1$, in Feynman gauge.}
\begin{ruledtabular}    
\begin{tabular}{|c|rrrrrrrrr|} 
$M$ & $N_s=8$ & $N_s=12$ & $N_s=16$ & $N_s=20$ & $N_s=24$ & $N_s=28$ 
    & $N_s=32$ & $N_s=48$ & $N_s=\infty$  
      \vspace{0.05cm} \\ \hline \vspace{-0.3cm} \\  
  0.1   &  0.11387 &  0.41183 &  0.64719 &  0.80302 &  0.89547 
        &  0.94650 &  0.97333 &  0.99859 &  1  \\
  0.2   &  0.71677 &  0.92503 &  0.98280 &  0.99634 &  0.99926 
        &  0.99985 &  0.99997 &  1.00000 &  1  \\
  0.3   &  0.94927 &  0.99550 &  0.99965 &  0.99997 &  1.00000 
        &  1.00000 &  1.00000 &  1.00000 &  1  \\
  0.4   &  0.99443 &  0.99986 &  1.00000 &  1.00000 &  1.00000 
        &  1.00000 &  1.00000 &  1.00000 &  1  \\
  0.5   &  0.99964 &  1.00000 &  1.00000 &  1.00000 &  1.00000
        &  1.00000 &  1.00000 &  1.00000 &  1  \\
  0.6   &  0.99999 &  1.00000 &  1.00000 &  1.00000 &  1.00000
        &  1.00000 &  1.00000 &  1.00000 &  1  \\
\hline    
\end{tabular}
\end{ruledtabular}
\end{table*}

\begin{table*}[htp]
\caption{\label{tab:div_vertex}
Coefficient of the logarithmic term of the vertex diagram of the vector 
current, in Feynman gauge.}
\begin{ruledtabular}    
\begin{tabular}{|c|rrrrrrrrr|} 
$M$ & $N_s=8$ & $N_s=12$ & $N_s=16$ & $N_s=20$ & $N_s=24$ & $N_s=28$ 
    & $N_s=32$ & $N_s=48$ & $N_s=\infty$  
      \vspace{0.05cm} \\ \hline \vspace{-0.3cm} \\  
  0.1   & -0.84792 & -1.02208 & -1.05864 & -1.04852 & -1.03113
        & -1.01779 & -1.00952 & -1.00058 & -1  \\
  0.2   & -1.04275 & -1.02123 & -1.00592 & -1.00138 & -1.00030
        & -1.00006 & -1.00001 & -1.00000 & -1  \\
  0.3   & -1.01354 & -1.00158 & -1.00014 & -1.00001 & -1.00000
        & -1.00000 & -1.00000 & -1.00000 & -1  \\
  0.4   & -1.00176 & -1.00005 & -1.00000 & -1.00000 & -1.00000
        & -1.00000 & -1.00000 & -1.00000 & -1  \\
  0.5   & -1.00012 & -1.00000 & -1.00000 & -1.00000 & -1.00000
        & -1.00000 & -1.00000 & -1.00000 & -1  \\
  0.6   & -1.00000 & -1.00000 & -1.00000 & -1.00000 & -1.00000
        & -1.00000 & -1.00000 & -1.00000 & -1  \\
\hline
\end{tabular}
\end{ruledtabular}
\end{table*}

\begin{table*}[htp]
\caption{\label{tab:div_vector}
Anomalous dimension of the vector current, in Feynman gauge.}
\begin{ruledtabular}    
\begin{tabular}{|c|rrrrrrrrr|} 
$M$ & $N_s=8$ & $N_s=12$ & $N_s=16$ & $N_s=20$ & $N_s=24$ & $N_s=28$ 
    & $N_s=32$ & $N_s=48$ & $N_s=\infty$  
      \vspace{0.05cm} \\ \hline \vspace{-0.3cm} \\  
  0.1   &  0.73405 &  0.61025 &  0.41144 &  0.24550 &  0.13567
        &  0.07128 &  0.03619 &  0.00199 &  0  \\ 
  0.2   &  0.32598 &  0.09620 &  0.02312 &  0.00504 &  0.00104
        &  0.00021 &  0.00004 &  0.00000 &  0  \\
  0.3   &  0.06427 &  0.00608 &  0.00049 &  0.00004 &  0.00000
        &  0.00000 &  0.00000 &  0.00000 &  0  \\
  0.4   &  0.00733 &  0.00020 &  0.00000 &  0.00000 &  0.00000
        &  0.00000 &  0.00000 &  0.00000 &  0  \\
  0.5   &  0.00048 &  0.00000 &  0.00000 &  0.00000 &  0.00000
        &  0.00000 &  0.00000 &  0.00000 &  0  \\
  0.6   &  0.00002 &  0.00000 &  0.00000 &  0.00000 &  0.00000
        &  0.00000 &  0.00000 &  0.00000 &  0  \\
\hline
\end{tabular}
\end{ruledtabular}
\end{table*}

\begin{table*}[htp]
\caption{\label{tab:div_scalar}
Anomalous dimension of the scalar density, in Feynman gauge.}
\begin{ruledtabular}    
\begin{tabular}{|c|rrrrrrrrr|} 
$M$ & $N_s=8$ & $N_s=12$ & $N_s=16$ & $N_s=20$ & $N_s=24$ & $N_s=28$ 
    & $N_s=32$ & $N_s=48$ & $N_s=\infty$  
      \vspace{0.05cm} \\ \hline \vspace{-0.3cm} \\  
  0.1   &  3.27780 &  3.67649 &  3.58735 &  3.39105 &  3.22907
        &  3.12464 &  3.06474 &  3.00372 &  3  \\
  0.2   &  3.45424 &  3.15990 &  3.04087 &  3.00918 &  3.00193
        &  3.00039 &  3.00008 &  3.00000 &  3  \\
  0.3   &  3.10489 &  3.01080 &  3.00089 &  3.00007 &  3.00000
        &  3.00000 &  3.00000 &  3.00000 &  3  \\
  0.4   &  3.01260 &  3.00036 &  3.00001 &  3.00000 &  3.00000
        &  3.00000 &  3.00000 &  3.00000 &  3  \\
  0.5   &  3.00085 &  3.00001 &  3.00000 &  3.00000 &  3.00000
        &  3.00000 &  3.00000 &  3.00000 &  3  \\
  0.6   &  3.00003 &  3.00000 &  3.00000 &  3.00000 &  3.00000
        &  3.00000 &  3.00000 &  3.00000 &  3  \\
\hline
\end{tabular}
\end{ruledtabular}
\end{table*}

\begin{table*}[htp]
\caption{\label{tab:sigma0tad}
Tadpole contribution to $\Sigma_0$, Eq.~(\ref{eq:tad-sigma0}), 
in Feynman gauge (multiplied for $16 \pi^2$).}
\begin{ruledtabular}    
\begin{tabular}{|c|rrrrrrrrr|} 
$M$ & $N_s=8$ & $N_s=12$ & $N_s=16$ & $N_s=20$ & $N_s=24$ & $N_s=28$ 
    & $N_s=32$ & $N_s=48$ & $N_s=\infty$  
      \vspace{0.05cm} \\ \hline \vspace{-0.3cm} \\  
  0.1   & 15.10361 & 16.98204 & 16.85601 & 15.31833 & 13.06260 
        & 10.63233 &  8.36032 &  2.59484 &  0  \\
  0.2   & 18.33890 & 12.98046 &  7.73201 &  4.17637 &  2.12604
        &  1.04115 &  0.49623 &  0.02182 &  0  \\ 
  0.3   & 12.61592 &  4.97585 &  1.66803 &  0.51427 &  0.15080 
        &  0.04277 &  0.01184 &  0.00006 &  0  \\
  0.4   &  6.03146 &  1.23559 &  0.21903 &  0.03602 &  0.00566 
        &  0.00086 &  0.00013 &  0.00000 &  0  \\
  0.5   &  2.10263 &  0.20309 &  0.01717 &  0.00135 &  0.00010 
        &  0.00001 &  0.00000 &  0.00000 &  0  \\ 
  0.6   &  0.51309 &  0.02003 &  0.00069 &  0.00002 &  0.00000 
        &  0.00000 &  0.00000 &  0.00000 &  0  \\
\hline    
\end{tabular}
\end{ruledtabular}
\end{table*}

\begin{table*}[htp]
\caption{\label{tab:sigma1tad}
Tadpole contribution to $\Sigma_1$, Eq.~(\ref{eq:tad-sigma1}), 
in Feynman gauge (multiplied for $16 \pi^2$).}
\begin{ruledtabular}    
\begin{tabular}{|c|rrrrrrrrr|} 
$M$ & $N_s=8$ & $N_s=12$ & $N_s=16$ & $N_s=20$ & $N_s=24$ & $N_s=28$ 
    & $N_s=32$ & $N_s=48$ & $N_s=\infty$  
      \vspace{0.05cm} \\ \hline \vspace{-0.3cm} \\  
  0.1   &  6.88684 &  8.89791 & 10.28165 & 11.14950 & 11.65586 
        & 11.93540 & 12.08339 & 12.22491 & 12.23305 \\
  0.2   & 10.71422 & 11.82437 & 12.13713 & 12.21230 & 12.22878 
        & 12.23220 & 12.23289 & 12.23305 & 12.23305 \\
  0.3   & 11.95987 & 12.20785 & 12.23105 & 12.23290 & 12.23304 
        & 12.23305 & 12.23305 & 12.23305 & 12.23305 \\
  0.4   & 12.20239 & 12.23224 & 12.23303 & 12.23305 & 12.23305 
        & 12.23305 & 12.23305 & 12.23305 & 12.23305 \\
  0.5   & 12.23104 & 12.23304 & 12.23305 & 12.23305 & 12.23305
        & 12.23305 & 12.23305 & 12.23305 & 12.23305 \\
  0.6   & 12.23299 & 12.23305 & 12.23305 & 12.23305 & 12.23305 
        & 12.23305 & 12.23305 & 12.23305 & 12.23305 \\
\hline
\end{tabular}
\end{ruledtabular}
\end{table*}

\begin{table*}[htp]
\caption{\label{tab:sigma0hc}
Coefficient of $\bar g^2$ for the contribution of the half-circle diagram 
to $\Sigma_0$, in Feynman gauge.}
\begin{ruledtabular}    
\begin{tabular}{|c|rrrrrrrrr|} 
$M$ & $N_s=8$ & $N_s=12$ & $N_s=16$ & $N_s=20$ & $N_s=24$ & $N_s=28$ 
    & $N_s=32$ & $N_s=48$ & $N_s=\infty$  
      \vspace{0.05cm} \\ \hline \vspace{-0.3cm} \\  
  0.1   &  2.75558  &  2.69676  &  2.39676  &  2.00277  &  1.60286 
        &  1.24317  &  0.94186  &  0.26825  & 0 \\
  0.2   &  3.48374  &  2.22615  &  1.24197  &  0.64382  &  0.31901
        &  0.15331  &  0.07208  &  0.00306  & 0 \\
  0.3   &  2.67348  &  1.01173  &  0.33194  &  0.10111  &  0.02942 
        &  0.00830  &  0.00227  &  0.00001  & 0 \\
  0.4   &  1.52701  &  0.31852  &  0.05710  &  0.00946  &  0.00147 
        &  0.00022  &  0.00003  &  0.00000  & 0 \\
  0.5   &  0.68927  &  0.07235  &  0.00641  &  0.00051  &  0.00004  
        &  0.00000  &  0.00000  &  0.00000  & 0 \\
  0.6   &  0.24943  &  0.01188  &  0.00046  &  0.00002  &  0.00000  
        &  0.00000  &  0.00000  &  0.00000  & 0 \\
\hline    
\end{tabular}
\end{ruledtabular}
\end{table*}

\begin{table*}[htp]
\caption{\label{tab:sigma0hccsi}
Coefficient of $\bar g^2$ for the part proportional to $1-\alpha$ of the 
contribution of the half-circle diagram to $\Sigma_0$.}
\begin{ruledtabular}    
\begin{tabular}{|c|rrrrrrrrr|} 
$M$ & $N_s=8$ & $N_s=12$ & $N_s=16$ & $N_s=20$ & $N_s=24$ & $N_s=28$ 
    & $N_s=32$ & $N_s=48$ & $N_s=\infty$  
      \vspace{0.05cm} \\ \hline \vspace{-0.3cm} \\  
  0.1   &  3.19742  &  3.77191  &  3.84954  &  3.56027  &  3.07177
        &  2.52079  &  1.99387  &  0.62668  & 0 \\ 
  0.2   &  4.00781  &  2.95012  &  1.79210  &  0.97875  &  0.50169
        &  0.24683  &  0.11804  &  0.00523  & 0 \\ 
  0.3   &  2.81125  &  1.14035  &  0.38726  &  0.12026  &  0.03543
        &  0.01008  &  0.00280  &  0.00001  & 0 \\ 
  0.4   &  1.35920  &  0.28390  &  0.05078  &  0.00839  &  0.00133
        &  0.00020  &  0.00003  &  0.00000  & 0 \\ 
  0.5   &  0.47745  &  0.04674  &  0.00398  &  0.00032  &  0.00002
        &  0.00000  &  0.00000  &  0.00000  & 0 \\ 
  0.6   &  0.11737  &  0.00463  &  0.00016  &  0.00001  &  0.00000
        &  0.00000  &  0.00000  &  0.00000  & 0 \\
\hline    
\end{tabular}
\end{ruledtabular}
\end{table*}

\begin{table*}[htp]
\caption{\label{tab:sigma0feynman}
Coefficient of $\bar g^2$ for the complete result of $\Sigma_0$, in Feynman 
gauge.}
\begin{ruledtabular}    
\begin{tabular}{|c|rrrrrrrrr|} 
$M$ & $N_s=8$ & $N_s=12$ & $N_s=16$ & $N_s=20$ & $N_s=24$ & $N_s=28$ 
    & $N_s=32$ & $N_s=48$ & $N_s=\infty$  
      \vspace{0.05cm} \\ \hline \vspace{-0.3cm} \\  
  0.1   & 17.85919  & 19.67880  & 19.25277  & 17.32109  & 14.66546
        & 11.87550  &  9.30218  &  2.86309  & 0  \\
  0.2   & 21.82264  & 15.20660  &  8.97397  &  4.82019  &  2.44504
        &  1.19446  &  0.56831  &  0.02488  & 0  \\
  0.3   & 15.28940  &  5.98758  &  1.99997  &  0.61538  &  0.18022
        &  0.05106  &  0.01412  &  0.00007  & 0  \\
  0.4   &  7.55847  &  1.55410  &  0.27613  &  0.04548  &  0.00713
        &  0.00108  &  0.00016  &  0.00000  & 0  \\
  0.5   &  2.79190  &  0.27543  &  0.02358  &  0.00186  &  0.00014
        &  0.00001  &  0.00000  &  0.00000  & 0  \\
  0.6   &  0.76252  &  0.03191  &  0.00115  &  0.00004  &  0.00000
        &  0.00000  &  0.00000  &  0.00000  & 0  \\
\hline    
\end{tabular}
\end{ruledtabular}
\end{table*}

\begin{table*}[htp]
\caption{\label{tab:sigma0landau}
Coefficient of $\bar g^2$ for the complete result of $\Sigma_0$, in Landau 
gauge.}
\begin{ruledtabular}    
\begin{tabular}{|c|rrrrrrrrr|} 
$M$ & $N_s=8$ & $N_s=12$ & $N_s=16$ & $N_s=20$ & $N_s=24$ & $N_s=28$ 
    & $N_s=32$ & $N_s=48$ & $N_s=\infty$  
      \vspace{0.05cm} \\ \hline \vspace{-0.3cm} \\  
  0.1   & 17.28070  & 19.20520  & 18.88831  & 17.05178  & 14.47158
        & 11.73820  &  9.20597  &  2.84106  & 0  \\
  0.2   & 21.24572  & 14.91161  &  8.83308  &  4.75484  &  2.41522
        &  1.18100  &  0.56229  &  0.02466  & 0  \\
  0.3   & 14.94668  &  5.88397  &  1.97022  &  0.60708  &  0.17795
        &  0.05045  &  0.01396  &  0.00007  & 0  \\
  0.4   &  7.40981  &  1.52910  &  0.27215  &  0.04487  &  0.00704
        &  0.00107  &  0.00016  &  0.00000  & 0  \\
  0.5   &  2.74369  &  0.27141  &  0.02327  &  0.00184  &  0.00014
        &  0.00001  &  0.00000  &  0.00000  & 0  \\
  0.6   &  0.75162  &  0.03153  &  0.00114  &  0.00004  &  0.00000
        &  0.00000  &  0.00000  &  0.00000  & 0  \\
\hline    
\end{tabular}
\end{ruledtabular}
\end{table*}

\section{Residual mass}
\label{sec:rm}

In this Section we report, for several choices of $N_s$ (and $M$), the results 
that we have obtained for $\Sigma_0$. This quantity enters into the description
of the one-loop quark self-energy, Eq.~(\ref{eq:seint}), and determines the 
critical (or residual) mass at this level. The numbers that we have obtained 
for $\Sigma_0$ are valid both in the quenched and unquenched cases, because 
at one loop internal quark loops can never appear in the diagrams that enter 
in this as well as in the other calculations presented in this paper. 
These diagrams are standard and well known and can be found for example 
in Ref.~\cite{Capitani:2005vb}. 

The behavior of the tadpole diagrams as $N_s$ and $M$ change is particularly 
interesting, and we think that it is useful to include here also the values 
of the tadpole contributing to $\Sigma_1$ (although in this work we need only 
the results of the tadpole contributing to $\Sigma_0$). 
Since no pure 5-dimensional quark propagators appear in the tadpoles, for these
diagrams the $\Sigma_{st} (p)$ of Eq.~(\ref{eq:sigmaq}) is diagonal in the 
fifth dimension and proportional to $(i\slash{k} - 4r/a) \, G_{\mu\nu}(k)$.
This is the same integrand of the tadpoles for Wilson fermions, and for that
action in the case of the tadpole diagram contributing to $\Sigma_1$ it 
gave the result (in a general covariant gauge) 
\begin{equation}
T_l = 8\pi^2 Z_0 \, (1-1/4\, (1-\alpha)) , 
\end{equation}
where $Z_0=0.154933390231\ldots$ is a well-known integral 
\cite{Capitani:2002mp}. It is then clear that for domain-wall fermions
(where we have now $r=-1$) the behavior of the tadpole diagrams as a function 
of $N_s$ (and $M$) is completely determined by the damping factors in the 
fifth dimension.
Their general effect can be seen by looking at their leading contributions 
for large $N_s$, which enter the game in the combinations
\begin{equation}
\sum_{s=1}^{N_s} w_0^{s-1} w_0^{N_s-s} = N_s w_0^{N_s-1} 
\end{equation}
and
\begin{equation}
\sum_{s=1}^{N_s} (w_0^2)^{s-1} = \sum_{s=1}^{N_s} (w_0^2)^{N_s-s} = 
\frac{1-w_0^{2N_s}}{1-w_0^2} .
\end{equation}
These are indeed the leading expressions, in units of
\begin{equation}
T_d = \frac{(1-w_0^2) \, T_l}{(1-w_0^{2N_s})^2}, 
\end{equation}
for the tadpole contributions to $\Sigma_0$ and $\Sigma_1$ respectively, 
in the limit of large $N_s$. From these expressions one can immediately see 
that the tadpole of $\Sigma_0$ vanishes when $N_s=\infty$, while the tadpole 
of $\Sigma_1$ gives in this limit the well-known Wilson number, $T_l$. 
The damping factors thus play a primary r\^ole in determining the results 
of the domain-wall tadpoles. We have calculated their exact expressions 
including all subleading terms in $N_s$, and the tadpole contribution 
to $\Sigma_0$ turns out to be equal to
\begin{equation}
  4 \, T_d \, \Big[ N_s \, (1 + w_0^{2(N_s+1)}) \, w_0^{N_s-1} 
       - 2 \, w_0^{N_s+1} \, \frac{1-w_0^{2N_s}}{1-w_0^2} \Big] ,
\label{eq:tad-sigma0}
\end{equation}
while the tadpole contribution to $\Sigma_1$ turns out to be equal to
\begin{equation}
T_d \, \Big[ (1 + w_0^{2(N_s+1)}) \, \frac{1-w_0^{2N_s}}{1-w_0^2} 
       - 2 \, N_s \, w_0^{2N_s} \Big] .
\label{eq:tad-sigma1}
\end{equation}
Numerical values of these tadpoles for various choices of $N_s$ and $M$ are 
collected in Tables \ref{tab:sigma0tad} and \ref{tab:sigma1tad}, where, as in 
the rest of the paper, we also show the corresponding values for the limiting 
case of infinite extent in the fifth dimension. In the case of $\Sigma_0$ we 
observe that its tadpole contribution presents wide variations with $N_s$ and 
$M$, so that in some cases it turns out to be small while in other cases it 
can be substantially large. The tadpole contributing to $\Sigma_1$ instead 
has smaller variations. This suggests that some care should be used when 
talking about tadpole dominance in relation to domain-wall fermions. 
In fact, the tadpoles contributing to $\Sigma_0$ and $\Sigma_1$ even decrease 
toward zero when $M\to 0$ or $M\to 2$, as we will see in 
Sect. \ref{sec:borders}. 

Given the strong dependence on $M$ and $N_s$, tadpole improvement (at least
in its more common form) seems in general not to be an appropriate tool with 
regard to the residual mass, which is here associated with the additive mass 
renormalization arising from the one-loop self-energy, Eq.~(\ref{eq:mres}). 
The contribution of the one-loop diagrams to this mass can be inferred from 
the numbers for $\Sigma_0$ presented in Tables \ref{tab:sigma0feynman} and 
\ref{tab:sigma0landau} (in Feynman and Landau gauge respectively), which take 
into account the results for the half-circle diagram given in Tables 
\ref{tab:sigma0hc} and \ref{tab:sigma0hccsi}.
Since we have performed the calculations in a general covariant gauge, we can 
express all the quantities presented in this paper in the form
\begin{equation}
A + (1-\alpha) \, B ,
\end{equation}
where $A$ and $A+B$ provide the answer in Feynman and Landau gauge 
respectively, and $B$ is a number which remains the same when using fermion 
formulations rather diverse like domain-wall with an infinite extent of the 
fifth dimension, Wilson or overlap. The values of $B$ in the case of the 
contribution of the half-circle diagram to $\Sigma_0$ at finite $N_s$ 
are given in Table \ref{tab:sigma0hccsi}. 
If we now compare the numbers shown in Tables \ref{tab:sigma0feynman} and 
\ref{tab:sigma0landau}, we can deduce that, even after the result for the 
tadpole is included, the contribution to $\Sigma_0$ proportional to $1-\alpha$ 
is in general not equal to zero, which is what one would have instead obtained 
if $N_s=\infty$ or if Wilson or overlap fermions were used. All this means 
that at finite $N_s$ the residual mass, which is derived from $\Sigma_0$ 
according to Eq.~(\ref{eq:mres}), is not a gauge invariant quantity anymore. 
Although numerically the deviations from gauge invariance remain in most cases 
rather small (because there are large cancellations between the contributions
of the half-circle and tadpole diagrams), we encounter here another of the 
pathological features of the domain-wall theory truncated at finite $N_s$. 
Thus, anomalous dimensions as well as terms proportional to $1-\alpha$, two 
of the quantities that remain the same when using a wide variety of fermionic 
actions, assume instead new values when the theory of domain-wall fermions
is truncated at a finite $N_s$. 

It is also interesting to compare the one-loop results presented here for 
$\Sigma_0$ to the numbers that one obtains using Wilson fermions, which are 
$-51.43471$ for the unimproved and $-31.98644$ for the improved 
(with $c_{sw}=1$) case. We can then see that the domain-wall results are 
much smaller than the Wilson numbers, even for $N_s$ as small as 8.

\begin{table*}[htp]
\caption{\label{tab:residual1}
Residual mass in lattice units at $\beta=6$ in Feynman gauge.} 
\begin{ruledtabular}    
\begin{tabular}{|c|rrrrrrrrr|} 
$M$ & $N_s=8$ & $N_s=12$ & $N_s=16$ & $N_s=20$ & $N_s=24$ & $N_s=28$ 
    & $N_s=32$ & $N_s=48$ & $N_s=\infty$  
      \vspace{0.05cm} \\ \hline \vspace{-0.3cm} \\  
  0.1   & -0.23258  & -0.21982  & -0.19777  & -0.16935  & -0.13898
        & -0.11021  & -0.08507  & -0.02538  & 0  \\
  0.2   & -0.24466  & -0.15313  & -0.08590  & -0.04485  & -0.02234
        & -0.01078  & -0.00508  & -0.00022  & 0  \\
  0.3   & -0.15850  & -0.05761  & -0.01858  & -0.00560  & -0.00162
        & -0.00045  & -0.00012  &  0.00000  & 0  \\
  0.4   & -0.07457  & -0.01452  & -0.00251  & -0.00041  & -0.00006
        & -0.00001  &  0.00000  &  0.00000  & 0  \\
  0.5   & -0.02650  & -0.00251  & -0.00021  & -0.00002  &  0.00000
        &  0.00000  &  0.00000  &  0.00000  & 0  \\
  0.6   & -0.00699  & -0.00028  & -0.00001  &  0.00000  &  0.00000
        &  0.00000  &  0.00000  &  0.00000  & 0  \\
\hline    
\end{tabular}
\end{ruledtabular}
\end{table*}

\begin{table*}[htp]
\caption{\label{tab:residual1l}
Residual mass in lattice units at $\beta=6$ in Landau gauge.} 
\begin{ruledtabular}    
\begin{tabular}{|c|rrrrrrrrr|} 
$M$ & $N_s=8$ & $N_s=12$ & $N_s=16$ & $N_s=20$ & $N_s=24$ & $N_s=28$ 
    & $N_s=32$ & $N_s=48$ & $N_s=\infty$  
      \vspace{0.05cm} \\ \hline \vspace{-0.3cm} \\  
  0.1   & -0.22770  & -0.21582  & -0.19469  & -0.16708  & -0.13735
        & -0.10905  & -0.08425  & -0.02520  & 0  \\
  0.2   & -0.23978  & -0.15064  & -0.08471  & -0.04430  & -0.02209
        & -0.01067  & -0.00503  & -0.00022  & 0  \\
  0.3   & -0.15560  & -0.05674  & -0.01833  & -0.00553  & -0.00160
        & -0.00045  & -0.00012  &  0.00000  & 0  \\
  0.4   & -0.07331  & -0.01430  & -0.00248  & -0.00040  & -0.00006
        & -0.00001  &  0.00000  &  0.00000  & 0  \\
  0.5   & -0.02610  & -0.00247  & -0.00021  & -0.00002  &  0.00000
        &  0.00000  &  0.00000  &  0.00000  & 0  \\
  0.6   & -0.00690  & -0.00028  & -0.00001  &  0.00000  &  0.00000
        &  0.00000  &  0.00000  &  0.00000  & 0  \\
\hline    
\end{tabular}
\end{ruledtabular}
\end{table*}

\begin{table*}[htp]
\caption{\label{tab:residual2}
Residual mass in lattice units at $\beta=5.2$ in Feynman gauge.} 
\begin{ruledtabular}    
\begin{tabular}{|c|rrrrrrrrr|} 
$M$ & $N_s=8$ & $N_s=12$ & $N_s=16$ & $N_s=20$ & $N_s=24$ & $N_s=28$ 
    & $N_s=32$ & $N_s=48$ & $N_s=\infty$  
      \vspace{0.05cm} \\ \hline \vspace{-0.3cm} \\  
  0.1   & -0.25578  & -0.24538  & -0.22278  & -0.19185  & -0.15803
        & -0.12564  & -0.09715  & -0.02910  & 0  \\
  0.2   & -0.27300  & -0.17289  & -0.09756  & -0.05111  & -0.02552
        & -0.01233  & -0.00582  & -0.00025  & 0  \\
  0.3   & -0.17836  & -0.06539  & -0.02118  & -0.00640  & -0.00185  
        & -0.00052  & -0.00014  &  0.00000  & 0  \\
  0.4   & -0.08439  & -0.01653  & -0.00287  & -0.00047  & -0.00007
        & -0.00001  &  0.00000  &  0.00000  & 0  \\
  0.5   & -0.03013  & -0.00287  & -0.00024  & -0.00002  &  0.00000
        &  0.00000  &  0.00000  &  0.00000  & 0  \\
  0.6   & -0.00798  & -0.00033  & -0.00001  &  0.00000  &  0.00000
        &  0.00000  &  0.00000  &  0.00000  & 0  \\
\hline    
\end{tabular}
\end{ruledtabular}
\end{table*}

\begin{table*}[htp]
\caption{\label{tab:residual2l}
Residual mass in lattice units at $\beta=5.2$ in Landau gauge.} 
\begin{ruledtabular}    
\begin{tabular}{|c|rrrrrrrrr|} 
$M$ & $N_s=8$ & $N_s=12$ & $N_s=16$ & $N_s=20$ & $N_s=24$ & $N_s=28$ 
    & $N_s=32$ & $N_s=48$ & $N_s=\infty$  
      \vspace{0.05cm} \\ \hline \vspace{-0.3cm} \\  
  0.1   & -0.25014  & -0.24077  & -0.21923  & -0.18923  & -0.15614
        & -0.12430  & -0.09621  & -0.02889  & 0  \\
  0.2   & -0.26738  & -0.17001  & -0.09619  & -0.05047  & -0.02523
        & -0.01220  & -0.00576  & -0.00025  & 0  \\
  0.3   & -0.17502  & -0.06438  & -0.02089  & -0.00632  & -0.00183
        & -0.00051  & -0.00014  &  0.00000  & 0  \\
  0.4   & -0.08294  & -0.01629  & -0.00283  & -0.00046  & -0.00007
        & -0.00001  &  0.00000  &  0.00000  & 0  \\
  0.5   & -0.02966  & -0.00283  & -0.00024  & -0.00002  &  0.00000
        &  0.00000  &  0.00000  &  0.00000  & 0  \\
  0.6   & -0.00787  & -0.00032  & -0.00001  &  0.00000  &  0.00000
        &  0.00000  &  0.00000  &  0.00000  & 0  \\
\hline    
\end{tabular}
\end{ruledtabular}
\end{table*}

\begin{table}[htp]
\caption{\label{tab:odd60}
Residual mass at $\beta=6.0$ for some odd $N_s$, in Landau gauge.} 
\begin{ruledtabular}    
\begin{tabular}{|c|rrrrrr|} 
   $M$ & $N_s=13$ &$N_s=14$ &$N_s=15$ &$N_s=16$ & $N_s=17$& $N_s=18$  
      \vspace{0.05cm} \\ \hline \vspace{-0.3cm} \\   
  0.1  &  -0.211  &  -0.206  &  -0.201  &  -0.195  &  -0.188  &  -0.181  \\
  0.2  &  -0.132  &  -0.114  &  -0.099  &  -0.085  &  -0.072  &  -0.062  \\
  0.3  &  -0.043  &  -0.033  &  -0.025  &  -0.018  &  -0.014  &  -0.010  \\
  0.4  &  -0.009  &  -0.006  &  -0.004  &  -0.002  &  -0.002  &  -0.001  \\
  0.5  &  -0.001  &  -0.001  &   0.000  &   0.000  &   0.000  &   0.000  \\
  0.6  &   0.000  &   0.000  &   0.000  &   0.000  &   0.000  &   0.000  \\
  0.7  &   0.000  &   0.000  &   0.000  &   0.000  &   0.000  &   0.000  \\
  0.8  &   0.000  &   0.000  &   0.000  &   0.000  &   0.000  &   0.000  \\
  0.9  &   0.000  &   0.000  &   0.000  &   0.000  &   0.000  &   0.000  \\
  1.0  &   0.000  &   0.000  &   0.000  &   0.000  &   0.000  &   0.000  \\
  1.1  &   0.000  &   0.000  &   0.000  &   0.000  &   0.000  &   0.000  \\
  1.2  &   0.000  &   0.000  &   0.000  &   0.000  &   0.000  &   0.000  \\
  1.3  &   0.000  &   0.000  &   0.000  &   0.000  &   0.000  &   0.000  \\
  1.4  &   0.000  &   0.000  &   0.000  &   0.000  &   0.000  &   0.000  \\
  1.5  &  -0.001  &   0.000  &   0.000  &   0.000  &   0.000  &   0.000  \\
  1.6  &  -0.006  &   0.004  &  -0.002  &   0.002  &  -0.001  &   0.001  \\
  1.7  &  -0.026  &   0.020  &  -0.015  &   0.012  &  -0.009  &   0.007  \\
  1.8  &  -0.074  &   0.067  &  -0.060  &   0.053  &  -0.047  &   0.041  \\
  1.9  &  -0.092  &   0.098  &  -0.102  &   0.105  &  -0.106  &   0.107  \\  
    \hline 
\end{tabular}
\end{ruledtabular}
\end{table}

\begin{table}[htp]
\caption{\label{tab:odd52}
Residual mass at $\beta=5.2$ for some odd $N_s$, in Landau gauge.} 
\begin{ruledtabular}    
\begin{tabular}{|c|rrrrrr|} 
   $M$ & $N_s=13$ &$N_s=14$ &$N_s=15$ &$N_s=16$ & $N_s=17$& $N_s=18$  
      \vspace{0.05cm} \\ \hline \vspace{-0.3cm} \\   
  0.1  &  -0.236  &  -0.231  &  -0.226  &  -0.219  &  -0.212  &  -0.205  \\
  0.2  &  -0.149  &  -0.129  &  -0.112  &  -0.096  &  -0.082  &  -0.070  \\
  0.3  &  -0.049  &  -0.037  &  -0.028  &  -0.021  &  -0.016  &  -0.012  \\
  0.4  &  -0.011  &  -0.007  &  -0.004  &  -0.003  &  -0.002  &  -0.001  \\
  0.5  &  -0.002  &  -0.001  &   0.000  &   0.000  &   0.000  &   0.000  \\
  0.6  &   0.000  &   0.000  &   0.000  &   0.000  &   0.000  &   0.000  \\
  0.7  &   0.000  &   0.000  &   0.000  &   0.000  &   0.000  &   0.000  \\
  0.8  &   0.000  &   0.000  &   0.000  &   0.000  &   0.000  &   0.000  \\
  0.9  &   0.000  &   0.000  &   0.000  &   0.000  &   0.000  &   0.000  \\
  1.0  &   0.000  &   0.000  &   0.000  &   0.000  &   0.000  &   0.000  \\
  1.1  &   0.000  &   0.000  &   0.000  &   0.000  &   0.000  &   0.000  \\
  1.2  &   0.000  &   0.000  &   0.000  &   0.000  &   0.000  &   0.000  \\
  1.3  &   0.000  &   0.000  &   0.000  &   0.000  &   0.000  &   0.000  \\
  1.4  &   0.000  &   0.000  &   0.000  &   0.000  &   0.000  &   0.000  \\
  1.5  &  -0.001  &   0.000  &   0.000  &   0.000  &   0.000  &   0.000  \\
  1.6  &  -0.006  &   0.004  &  -0.003  &   0.002  &  -0.001  &   0.001  \\
  1.7  &  -0.030  &   0.024  &  -0.018  &   0.014  &  -0.010  &   0.008  \\
  1.8  &  -0.089  &   0.080  &  -0.071  &   0.063  &  -0.055  &   0.048  \\
  1.9  &  -0.114  &   0.120  &  -0.124  &   0.126  &  -0.127  &   0.127  \\ 
    \hline
\end{tabular}
\end{ruledtabular}
\end{table}

\begin{table*}[htp]
\caption{\label{tab:va}
Coefficient of $\bar g^2$ for the quantity $\Delta = Z_V - Z_A = -(Z_S-Z_P)/2$,
which vanishes when chiral symmetry is restored at $N_s=\infty$, in Feynman 
gauge.}
\begin{ruledtabular}    
\begin{tabular}{|c|rrrrrrrrr|} 
$M$ & $N_s=8$ & $N_s=12$ & $N_s=16$ & $N_s=20$ & $N_s=24$ & $N_s=28$ 
    & $N_s=32$ & $N_s=48$ & $N_s=\infty$  
      \vspace{0.05cm} \\ \hline \vspace{-0.3cm} \\  
  0.1   &  6.9399  &  4.4434  &  2.6941  &  1.5749  &  0.8916  
        &  0.4899  &  0.2622  &  0.0179  & 0 \\
  0.2   &  2.3272  &  0.7211  &  0.1945  &  0.0477  &  0.0110 
        &  0.0025  &  0.0005  &  0.0000  & 0 \\
  0.3   &  0.5894  &  0.0668  &  0.0064  &  0.0006  &  0.0000  
        &  0.0000  &  0.0000  &  0.0000  & 0 \\
  0.4   &  0.0994  &  0.0034  &  0.0001  &  0.0000  &  0.0000
        &  0.0000  &  0.0000  &  0.0000  & 0 \\
  0.5   &  0.0106  &  0.0001  &  0.0000  &  0.0000  &  0.0000
        &  0.0000  &  0.0000  &  0.0000  & 0 \\
  0.6   &  0.0007  &  0.0000  &  0.0000  &  0.0000  &  0.0000
        &  0.0000  &  0.0000  &  0.0000  & 0 \\
\hline
\end{tabular}
\end{ruledtabular}
\end{table*}

\begin{table*}[htp]
\caption{\label{tab:vacsi}
Coefficient of $\bar g^2$ for the part proportional to $1-\alpha$ of $\Delta$.}
\begin{ruledtabular}    
\begin{tabular}{|c|rrrrrrrrr|} 
$M$ & $N_s=8$ & $N_s=12$ & $N_s=16$ & $N_s=20$ & $N_s=24$ & $N_s=28$ 
    & $N_s=32$ & $N_s=48$ & $N_s=\infty$  
      \vspace{0.05cm} \\ \hline \vspace{-0.3cm} \\  
  0.1   & -1.1115  & -0.8058  & -0.5284  & -0.3257  & -0.1914
        & -0.1081  & -0.0590  & -0.0042  & 0 \\
  0.2   & -0.4075  & -0.1438  & -0.0416  & -0.0106  & -0.0025
        & -0.0006  & -0.0001  &  0.0000  & 0 \\
  0.3   & -0.1092  & -0.0140  & -0.0014  & -0.0001  &  0.0000
        &  0.0000  &  0.0000  &  0.0000  & 0 \\
  0.4   & -0.0191  & -0.0007  &  0.0000  &  0.0000  &  0.0000
        &  0.0000  &  0.0000  &  0.0000  & 0 \\
  0.5   & -0.0020  &  0.0000  &  0.0000  &  0.0000  &  0.0000
        &  0.0000  &  0.0000  &  0.0000  & 0 \\
  0.6   & -0.0001  &  0.0000  &  0.0000  &  0.0000  &  0.0000
        &  0.0000  &  0.0000  &  0.0000  & 0 \\
\hline
\end{tabular}
\end{ruledtabular}
\end{table*}

\begin{table*}[htp]
\caption{\label{tab:d1mix}
Coefficient of $\bar g^2$ for the coefficient of the power-divergent mixing 
of $O_{d_1}$, $c_{mix}$ in Eq.~(\ref{eq:g2mixing}), in Feynman gauge.}
\begin{ruledtabular}    
\begin{tabular}{|c|rrrrrrrrr|} 
$M$ & $N_s=8$ & $N_s=12$ & $N_s=16$ & $N_s=20$ & $N_s=24$ & $N_s=28$ 
    & $N_s=32$ & $N_s=48$ & $N_s=\infty$  
      \vspace{0.05cm} \\ \hline \vspace{-0.3cm} \\  
  0.1   &-14.1244  & -8.4448  & -5.3682  & -3.4884  & -2.2846
        & -1.4996  & -0.9848  & -0.1829  & 0 \\
  0.2   & -5.8166  & -2.4014  & -0.9910  & -0.4075  & -0.1674
        & -0.0688  & -0.0283  & -0.0008  & 0 \\
  0.3   & -2.4733  & -0.6070  & -0.1475  & -0.0359  & -0.0087
        & -0.0021  & -0.0005  &  0.0000  & 0 \\
  0.4   & -0.9361  & -0.1265  & -0.0169  & -0.0022  & -0.0003
        &  0.0000  &  0.0000  &  0.0000  & 0 \\
  0.5   & -0.3094  & -0.0213  & -0.0014  & -0.0001  &  0.0000
        &  0.0000  &  0.0000  &  0.0000  & 0 \\
  0.6   & -0.0893  & -0.0029  & -0.0001  &  0.0000  &  0.0000
        &  0.0000  &  0.0000  &  0.0000  & 0 \\
\hline
\end{tabular}
\end{ruledtabular}
\end{table*}

\begin{table*}[htp]
\caption{\label{tab:d1mixcsi}
Coefficient of $\bar g^2$ for the part proportional to $1-\alpha$ 
of $c_{mix}$.}
\begin{ruledtabular}    
\begin{tabular}{|c|rrrrrrrrr|} 
$M$ & $N_s=8$ & $N_s=12$ & $N_s=16$ & $N_s=20$ & $N_s=24$ & $N_s=28$ 
    & $N_s=32$ & $N_s=48$ & $N_s=\infty$  
      \vspace{0.05cm} \\ \hline \vspace{-0.3cm} \\  
  0.1   &  0.3241  &  0.1056  &  0.0395  &  0.0162  &  0.0071
        &  0.0033  &  0.0015  &  0.0001  & 0 \\
  0.2   &  0.0480  &  0.0093  &  0.0019  &  0.0004  &  0.0001
        &  0.0000  &  0.0000  &  0.0000  & 0 \\
  0.3   &  0.0093  &  0.0008  &  0.0001  &  0.0000  &  0.0000  
        &  0.0000  &  0.0000  &  0.0000  & 0 \\
  0.4   &  0.0014  &  0.0000  &  0.0000  &  0.0000  &  0.0000
        &  0.0000  &  0.0000  &  0.0000  & 0 \\
  0.5   &  0.0001  &  0.0000  &  0.0000  &  0.0000  &  0.0000
        &  0.0000  &  0.0000  &  0.0000  & 0 \\
  0.6   &  0.0000  &  0.0000  &  0.0000  &  0.0000  &  0.0000
        &  0.0000  &  0.0000  &  0.0000  & 0 \\
\hline
\end{tabular}
\end{ruledtabular}
\end{table*}

The values of $a m_{res}^{(1)}$ which come out from our results for $\Sigma_0$
are reported in Tables \ref{tab:residual1} to \ref{tab:residual2l}.
One can immediately notice that the deviations from the case of exact chiral 
symmetry are much more pronounced when $M$ is close to 0.1 or 1.9. Since 
$M=1.8$ is at present a preferred choice for Monte Carlo simulations, as it 
appears to minimize chiral violations at the nonperturbative level 
\cite{Blum:1996jf,Blum:1997mz}, we can focus on this value of $M$ and observe 
that at one loop the critical mass seems to be still large up to $N_s=16$ or
$N_s=20$, becoming then somewhat smaller for $N_s=24$ and higher. 
Since the tree level and one-loop contributions are of a different sign, it 
could also be that the two-loop expression enters again with a negative sign 
and that its effect is to damp the chiral violations that we have obtained. 
But if higher-order corrections do not at the end strongly compensate 
the results of the residual mass at one loop, all this suggests that 
values like $N_s=24$ would be better choices for the simulations.

The region around $M=1.8$ suffers perturbatively rather badly from another 
problem as well. When $M$ is indeed close to 0.1 or 1.9, the numerical 
convergence of the integrals (which are computationally quite demanding, 
being in this case sums over a six-dimensional space) is much worse then 
when $M$ is nearer to 1. This behavior had already been noticed in the 
calculations for the $N_s=\infty$ limit presented in \cite{Capitani:2005vb}, 
however now the convergence is even slower than in that case, and it also 
tends to become worse when $N_s$ is small. 

The values of the residual mass $a m_{res}^{(1)}$, that is after the one-loop 
corrections are added to the tree-level expression, have the same sign of the 
critical mass of Wilson fermions only for $M \gtrsim 1.2$. In this region of
$M$ they are positive, at least for even $N_s$ and if the coupling is not very 
small. If one looks at the columns corresponding to $N_s=8$, our results seem 
to indicate that the minimal amount of chiral violations is attained for 
$M \sim 1.2$ (in Feynman or Landau gauge). The optimal choice of $M$ from 
the point of view of 1-loop calculations would then be around these values. 
This effect can be related to the renormalization of $M$, a quantity which 
is not protected by chiral symmetry and is moved by radiative corrections away 
from its free value, $M=1$ (where the tree-level residual mass vanishes). 
One can conjecture that higher-loop corrections and nonperturbative effects 
would shift this optimal value further on, until the chiral violations
approach a minimal point around $M=1.8$, which seems to be the choice that 
provides the smallest residual mass in Monte Carlo simulations. 
It could be that the results of the one-loop calculations for $M=1.8$
are not substantially changed once higher-loop corrections are included, 
while for $M<1.8$ they are increased by renormalization so that the minimum 
finally ends up located near $M=1.8$.

Notice that for $M=1.9$ the residual mass at $N_s=12$ is larger than at 
$N_s=8$, and at $N_s=16$ is even larger. We will discuss more in detail 
this kind of behavior in Sect. \ref{sec:borders}.

As an aside, we have also investigated what happens when one chooses an odd 
value of $N_s$. As can be seen from Tables \ref{tab:odd60} and \ref{tab:odd52},
which are representative of the general situation, when $M$ is smaller than 1 
the residual mass has always a negative sign for $N_s$ even or odd, but in 
the more interesting case of $M$ greater than 1 the one-loop residual mass
turns out to be positive when $N_s$ is even and negative when $N_s$ is odd. 
This could be related to the suggestion which was made by Neuberger some years 
ago \cite{Neuberger:1997bg}, according to which for odd $N_s$ the theory which 
is simulated could correspond to the $\theta=\pi$ regime of QCD. 

The residual mass obviously changes also when the coupling $g_0$ is varied, 
as an effect of the loop corrections. Indeed, the overlap between the chiral 
modes living near the two walls depends also on the strength of the gauge 
coupling, and for strong couplings this overlap tends to acquire some 
nonnegligible values. Furthermore, the residual mass explicitly depends also on
the value of the lattice spacing $a$, as can be seen from Eq.~(\ref{eq:mres}), 
since all the numbers presented in the Tables are given, as elsewhere in the 
paper, in lattice units. The residual mass is thus different for quenched and 
unquenched simulations made at the same lattice spacing, because $a$ and $g_0$ 
are related in a different way. Let us now consider some typical values of the 
parameters at which simulations are currently performed. We first fix $M=1.8$ 
and $N_s=16$. In the quenched case we can then take $\beta=6.0$, which 
corresponds to a lattice spacing of about 2 GeV, and according to our one-loop 
calculations this gives $a m_{res} = 0.05207$ in Feynman gauge and 
$a m_{res} = 0.05326$ in Landau gauge, that is $m_{res} \sim 104$ MeV and 
$m_{res} \sim 107$ MeV respectively. The dependence on the gauge seems thus 
from the numerical point of view not to be very significant. Notice also that 
if $m_{res}^{(0)}$ in Eq. \ref{eq:mres_tree} were positive, the values of 
$m_{res}^{(1)}$ would be even larger.
In the unquenched case, if we take $\beta=5.2$, which corresponds roughly to 
the same lattice spacing of 2 GeV but now to a different bare coupling, we 
then obtain $a m_{res} = 0.06164$ in Feynman gauge and $a m_{res} = 0.06302$
in Landau gauge, that is $m_{res} \sim 123$ MeV and $m_{res} \sim 126$ MeV 
respectively. Results from dynamical domain-wall simulations show that the 
residual mass in full QCD assumes much higher values than in the quenched case.
Dynamical domain-wall fermions present then a larger explicit chiral symmetry 
breaking. We can infer a similar effect which goes in the same direction from 
our one-loop perturbative results, although here the difference between
quenched and full QCD is not as pronounced as it has been observed
in the simulations, suggesting that higher-order corrections and especially 
nonperturbative effects presumably play a significant r\^ole.

We have automated the calculations presented in this article by developing 
suitable FORM codes \cite{Vermaseren:2000nd}, which are a sizeable extension 
of the programs which were used in the calculations presented in 
Ref. \cite{Capitani:2005vb} for $N_s=\infty$. 
These codes are now able to compute matrix elements for general values 
of $N_s$ and $M$. 

The integrals of the Feynman diagrams have been numerically evaluated by 
keeping $N_s$ fixed while at the same time refining the integration grid 
in the usual four-dimensional momentum space (where we needed to consider 
values of $L_x=120$). One might wonder whether this way of performing the
computation of the integrals, which amounts to taking a finite extra dimension 
with a small number of points while the number of points in the other four 
dimensions goes to infinity, is legitimate. One could after all object that 
the number of lattice points in the standard four dimensions is also bound
to be finite in any Monte Carlo simulation (although it can reach the order 
of $10^2$ for each of the coordinates, a number certainly larger than any 
at present practical value of $N_s$).

We can however observe that in the usual four-dimensional momentum space one 
does not encounter so wide variations for the Feynman integrand functions 
as instead is the case in the fifth dimension, where the Feynman integrands 
contain functions with an exponential behavior, responsible for the fact that 
for small $N_s$ (of the order of ten) large deviations of the integrals from 
their asymptotic values can be observed. Indeed, refining the integration grid 
in the usual four-dimensional momentum space from, say, $L_x=60$ to $L_x=80$, 
generally produces minimal differences in the momentum integrals, and it is 
only in order to extract five or more significant digits that one needs to use 
$L_x=100$ or higher. 
I think that it is important in this regard to keep in mind that the shape of 
the propagator in the fifth dimension corresponds to physical effects strictly 
connected with chirality (or lack thereof), whereas in 4-dimensional momentum 
space it corresponds to finite size effects only. One could also take the point
of view in which a lattice spacing $a_s$ for the fifth dimension, which is 
distinct from the four-dimensional lattice spacing $a$, is introduced, 
and imagine a situation where one keeps $a_s$ finite while the usual $a$ 
goes to zero.

In the domain-wall perturbative calculations for infinite $N_s$ made so far 
it was implicitly assumed that one could take the limit $N_s\to\infty$ for 
the quark propagators before actually using them to perform the computation
of the Feynman diagrams and the momentum integration. This could in principle 
present some problems, because these asymptotic propagators know about the 
Atiyah-Singer index, which is something which is not present in the theory 
at finite $N_s$, and furthermore they possess extra infrared singularities 
because of the exactly massless chiral mode. In the present work we always 
in the first place compute the Feynman diagrams and the integrals using the 
exact Feynman rules appropriate for finite $N_s$, and only afterwards we try
to investigate the limit $N_s \to \infty$ by computing the integrals anew for 
each increasing value of $N_s$. We can then observe that, apart from extreme 
values of $M$ (see Sect. \ref{sec:borders}), these integrals rapidly approach 
the values obtained with the theory which uses the $N_s=\infty$ propagators. 
By using the exact theory with finite $N_s$, we have hence also provided the 
check that in the large $N_s$ limit one indeed recovers the results obtained 
using the simpler asymptotic theory. We can thus confirm, at least numerically,
that the inversion of the limits was legitimate. 

In the case of the contribution of the half-circle diagram of the self-energy
to $\Sigma_0$, we have also repeated the whole calculation (in a general 
covariant gauge) by hand, including the evaluation of the gamma algebra and the
explicit computation of the sums over the fifth-dimensional indices. The final 
expressions are very lengthy and it is of no interest to report them here, 
however this alternative procedure provides a rather independent check of our 
calculations with respect to both the analytic part and the precision of the 
numerical integration in the 6-dimensional code. Furthermore, since this time 
we compute the sums in the extra dimension analytically, this saves 2 
dimensions in the numerical integration, which is reduced to the usual 4, 
and this is why we are able to give the results for $\Sigma_0$ with more 
precision than otherwise. Of course we cannot always employ this procedure, 
given that the other diagrams are in general way too complicated for a 
computation by hand. But this alternative 4-dimensional code for $\Sigma_0$ 
provides another advantage because the dependence on $N_s$ is now computed 
exactly and the computational cost is the same for every $N_s$, and thus 
we can think of getting results also at higher values of $N_s$, which would 
be too expensive in the 6-dimensional code where the computational cost 
grows as $N_s^2$. 

We conclude this Section by reminding that the tadpoles of course can be
calculated for any $N_s$ and $M$ with an extremely high precision, which is 
limited only by the knowledge of $Z_0$ (at present known with about 400
digits \cite{Capitani:2002mp}).

\section{Bilinear differences}
\label{sec:va}

We now consider the calculation at finite $N_s$ of matrix elements of some 
operators. In this Section we present one-loop results for the (finite) 
differences of chirally-related bilinear operators, which should become zero 
at infinite $N_s$, that is when chiral symmetry is fully restored. Since the 
vector and axial-vector currents renormalize differently when chiral symmetry 
is broken, an estimate of chirality-breaking effects can indeed be given 
by how much for a given finite $N_s$ the perturbative results for these 
currents differ from each other. The quantity $\Delta = Z_V - Z_A$ provides 
such a measure of chirality breaking. Moreover, one finds that 
$Z_V - Z_A = -(Z_S - Z_P)/2$ \cite{Duerr}.
The fact that we obtain for $\Delta$ the same number with a very good 
precision whether we consider the vector and axial-vector case or the scalar 
and pseudoscalar case then provides a compelling check of our calculations.   

We can successfully obtain all results for $\Delta$ by computing only 
finite lattice diagrams, because of the following three facts. For one thing, 
the anomalous dimensions of the operators which are chirally related are 
the same also when one considers the subleading orders (see Sect. 
\ref{sec:divergences}), that is $\gamma_{V}^{(N_s,M)} = \gamma_{A}^{(N_s,M)}$ 
and $\gamma_{S}^{(N_s,M)} = \gamma_{P}^{(N_s,M)}$. Furthermore, the 
continuum values of the finite parts, called $R^{\overline{\mathrm{MS}}}$ in 
Eq.~(\ref{eq:1loopcont}), are also equal for these pairs of operators. 
Finally, the mismatch between lattice (at finite $N_s$) and continuum anomalous
dimensions that we discussed at the end of Sect. \ref{sec:divergences} is 
also the same for the chirally-related operators, and thus it cancels in their 
differences. We remind that the vector and axial-vector currents have now a 
nonvanishing anomalous dimension, which becomes zero only when $N_s=\infty$. 

As we can gather in Tables \ref{tab:va} and \ref{tab:vacsi}, the decrease of 
the amount of chirality breaking connected to $\Delta$ follows a pattern 
similar to the one that we have encountered in the case of the critical mass 
and to the one that we will see for the operator discussed in the next 
Section, that is $\Delta$ is rather large for small $N_s$ and large $|1-M|$, 
and decreases when $N_s$ grows or when $|1-M|$ tends towards zero. Again, 
the part proportional to $(1-\alpha)$ of $\Delta$ shows a violation of gauge 
invariance, since if gauge invariance were respected these numbers would have 
to be zero, as it happens for Wilson or overlap fermions. We can notice that 
here the deviations from gauge invariance are of the same order as the ones 
for the critical mass, and that they are small. The operator studied in the 
next Section gives also very similar results regarding this point.

Taking a typical case, where $M=1.8$ and $N_s=16$ and $a^{-1} \sim 2$~GeV,
the numbers that we have obtained imply for QCD a chiral violation of about 
2 MeV, with the quenched case giving the slightly lower value. These numbers 
are much smaller, at given $M$ and $N_s$, than the ones obtained for $m_{res}$.
This could mean that the chiral violations are here of $O(m_{res}^2)$, as it 
has been suggested for other quantities in \cite{Aoki:2004ht,Christ:2005xh}.

It is also interesting to compare the numbers presented in this Section to 
the results that one obtains for this quantity when Wilson fermions are used. 
In this case one gets $\Delta = 4.82152$ for the unimproved case and 
$\Delta = 1.53633$ for the improved case, which correspond to chiral violations
which are at least of order 30 MeV for $\beta \sim 6.0$.

\section{A power-divergent mixing}
\label{sec:d1}

We think that it is also very instructive to study the case of an operator 
mixing which gets completely suppressed only when chiral symmetry is fully 
operative, so that the nonzero amount of residual mixing present for any given 
finite $N_s$ provides another quantitative measure of chiral violations. 
One of the simplest examples of this kind of mixings is probably furnished
by the operator
\begin{equation}
O_{d_1} = \bar q \gamma_{[4} \gamma_5 D_{1]} q ,
\end{equation}
which taken together with $O_{a_2,d} = 
\bar q \gamma_{\{1} \gamma_5 D_{4\}} q$ (or, in an alternative representation
of the hypercubic group, $O_{a_2,e} = \bar q \gamma_4 \gamma_5 D_4 q
            -\frac{1}{3} \sum_{i=1}^3  \bar q \gamma_i \gamma_5 D_i q$),
determines the first moment of the $g_2$ structure function, which measures 
the distribution of the (chiral even) transverse spin of quarks inside hadrons,
and also receives contributions from twist-3 operators. More details on these 
operators can be found in \cite{Capitani:2000wi,Capitani:2000aq} (of which 
we follow the notation) and references therein. Matrix elements of the operator
$O_{d_1}$ have been recently simulated using quenched domain-wall fermions 
with the DBW2 gauge action \cite{Orginos:2005uy}.
We remind that the symbol $[]$ denotes antisymmetrization over the relevant 
Lorentz indices, while $\{\}$ denotes symmetrization, and that for the 
covariant derivatives $\stackrel{\phantom{\rightarrow}}{D}=
\stackrel{\rightarrow}{D}-\stackrel{\leftarrow}{D}$
we use the lattice discretizations
\begin{equation}
\stackrel{\rightarrow}{D_\mu} q (x) = \frac{1}{2} \, \Big[U_\mu (x)
q(x+\hat{\mu}) -U_\mu^\dagger (x-\hat{\mu}) q(x-\hat{\mu}) \Big]
\end{equation}
\begin{equation}
\bar q (x) \stackrel{\leftarrow}{D_\mu} = \frac{1}{2} \, \Big[
\bar q (x+\hat{\mu}) U_\mu^\dagger (x) - \bar q(x-\hat{\mu})
U_\mu (x-\hat{\mu}) \Big] .
\end{equation}

We point out that many of the mixings which one encounters in the study 
of physical processes, like for instance the ones occurring in the 
renormalization of the second moment of the unpolarized parton distribution
\cite{Beccarini:1995iv,Gockeler:1995wg}, arise as a consequence of the 
breaking of Lorentz symmetry (or, in other cases, of other symmetries apart 
from chirality) and hence do not interest us much in the present context. 
The power-divergent mixing on the lattice of $O_{d_1}$ with an operator 
of lower dimension, which can be written as
\begin{equation}
c_{mix} \cdot \frac{i}{a} \, \bar q \sigma_{41} \gamma_5 q ,
\label{eq:g2mixing}
\end{equation}
is instead particularly interesting because is only caused by the breaking 
of chirality, and hence it provides a quantitative measure of how much 
chiral symmetry has been broken. The finite coefficient $c_{mix}$ must vanish 
for infinite $N_s$, when chiral symmetry is fully restored, and $O_{d_1}$ 
becomes in this case multiplicatively renormalized (like it also is when 
overlap fermions are used). For finite $N_s$ the values of this coefficient 
in the Feynman gauge are reported in Table \ref{tab:d1mix}, where we can see 
that in general this mixing can be considered to be almost negligible. 
Were this not the case, the removal of these lattice artifacts in Monte Carlo 
simulations would become quite challenging.

The results show that large chiral violations are present only for very small
$N_s$, or when $M$ is rather close to 0 or 2. Taking a typical case, for 
$M=1.8$ and $N_s=16$, and $a^{-1} \sim 2$~GeV, the chiral violations come out 
of about 3 MeV, with the quenched case giving the slightly lower value. 
These numbers are much smaller, at given $M$ and $N_s$, than the ones obtained 
for $m_{res}$. It could be that, as it has been suggested for other quantities 
in \cite{Aoki:2004ht,Christ:2005xh}, the chiral violations are here quadratic 
in $m_{res}$, that is this coefficient is doubly suppressed.

Again, as can be deduced from Table \ref{tab:d1mixcsi}, gauge invariance is 
lost in the theory at finite $N_s$. The part proportional to $1-\alpha$ of 
$c_{mix}$ vanishes only when $N_s=\infty$ (and this happens also when for 
example Wilson or overlap fermions are used). 

Wilson fermions also suffer from this power-divergent mixing which is caused
by the breaking of chirality, but the mixing coefficient is in this case gauge 
invariant. However, it takes the values $c_{mix} = 16.243762$ for the 
unimproved and $c_{mix} = 8.798732$ for the improved case with $c_{sw}=1$,
so that compared to domain-wall fermions at the standard choices of $N_s$ and 
$M$, the Wilson violations are then about two orders of magnitude higher. 
For overlap fermions of course $c_{mix}$ is zero.

\section{Toward the borders}
\label{sec:borders}

It it also interesting to see what happens when the domain-wall height $M$ 
is chosen dangerously close to 0 or 2, that is at the edges of its (at lowest 
order) allowed values. After investigating in various cases, we have found 
as a general phenomenon that the exponential decay in $N_s$ of the chiral 
violations slows down when approaching these borders. In some instances it 
can even happen that for a fixed $M$ chosen very near to one of the edges 
the chiral violations increase with $N_s$, at least up to a certain point.
 
One can observe this behavior already by looking at the tadpole diagrams, 
which can be calculated exactly. For $M\to 0$ they assume in fact the 
asymptotic expressions
(in Feynman gauge)
\begin{equation}
\frac{4}{3} \, M \, T_d \left( N_s +3 +\frac{2}{N_s} \right) 
\label{eq:limtad0}
\end{equation}
in the case of $\Sigma_0$, and 
\begin{equation}
\frac{1}{3} \, M \, T_d \left( 2 N_s +3 +\frac{1}{N_s}  \right)
\label{eq:limtad1}
\end{equation}
in the case of $\Sigma_1$. For $M\to 2$ one just takes into account the fact 
that these tadpoles are respectively odd and even upon reflection around 
the point $M=1$.
We can then immediately see that for a given $N_s$ the tadpoles tend to zero 
when $M$ approaches 0 or 2. From Table \ref{tab:sigma0tad} we can also see 
that the decay in $N_s$ (at a constant $M$) of the tadpole of $\Sigma_0$ tends 
to become slower and slower as $M$ nears the borders, until the rate of decay 
probably vanishes at some point and the tadpole then reaches the asymptotic 
regime of Eq. (\ref{eq:limtad0}), where actually the chiral violations grow 
with $N_s$ (at fixed $M$). We can observe a similar behavior also for the 
other quantities considered in this article, which we have run for values 
of $M$ close to $|1-M| \sim 0.99$ and $|1-M| \sim 0.999$. 
For $\Delta$ and $c_{mix}$ the rate of the exponential decay in $N_s$ keeps
decreasing when one approaches the borders, until a likely final disappearance,
even though these quantities instead increase in $M$ (for a fixed $N_s$) when 
$M\to 0$ and $M\to 2$. In fact, $c_{mix}$ tends to diverge very fast when 
$M\to 0$.

The slowing of the exponential decays of $m_{res}$, $\Delta$ and $c_{mix}$ when
one moves $M$ such that $|1-M|\to 1$ can be related to the decrease of the 
mobility edge $\lambda_c$ towards zero in these extreme regions of $M$ 
\cite{Golterman:2003qe,Golterman:2004cy,Golterman:2005fe}. 
Choosing a value of $M$ too close to 0 or 2 can then become dangerous, because 
the mobility edge has to remain well above zero in order to perform reliable 
Monte Carlo simulations, otherwise the restoration of chiral symmetry can 
become problematic. The fall of the mobility edge to zero signals the onset 
of the Aoki phase, and this can be pictorially seen for example in Fig.~1 of 
Ref.~\cite{Golterman:2003qe} or \cite{Golterman:2005fe}, where in fact it 
corresponds to moving, for $g_0$ not too large, from the rightmost $C$ phase
towards the $B$ phase, which one eventually enters through one of the 
thin ``fingers''.

We can thus briefly sum up the behavior of the chiral violations when $M$ 
changes as follows. The coefficient of their exponential decay is zero or 
rather small when $M$ is in the vicinity of 1. It then grows when $M$ moves 
towards either 0 or 2, before decreasing again and at last getting rather 
small for $M\sim 0$ and $M\sim 2$. 

A remarkable thing that we have observed for values of $M$ very near the edges
is that $m_{res}$ and $\Delta$ happen to be smaller for small $N_s$ than for 
larger values of $N_s$, that is the chiral violation initially grow in $N_s$ 
instead of decaying exponentially. This observed initial growth is just a 
temporary one, before eventually $m_{res}$ or $\Delta$ starts to decay as 
expected. But it could be that eventually a behavior like the one of 
Eqs. (\ref{eq:limtad0}) and (\ref{eq:limtad1}) sets in, that is the exponential
decay disappears altogether and one can only see a steady increase with $N_s$.

For $c_{mix}$ this does not seem to happen, that is we always observe a clear 
exponential-like decay, without any initial growth in $N_s$, at least for 
$N_s \ge 8$ and for the values of $M$ that we have investigated, up to 
$2-M=10^{-10}$. 

Numerical examples that illustrate this phenomenon of the initial growth 
in $N_s$ can be seen in the case of $m_{res}$ in Tables 
\ref{tab:limitmresfeynman} and \ref{tab:limitmreslandau}, which refer 
to $M\to 2$ (similar things can also be seen for the other limit, $M\to 0$). 
The onset of this behavior takes place around $M=1.9$, as can also be observed
in the last line of Tables \ref{tab:residual1} to \ref{tab:residual2l}.
Furthermore, we can notice that the exponential decay which follows these 
regions of initial growth in $N_s$ sets in at ever higher values of $N_s$ 
when one gets nearer and nearer to 2 (note that for $M>1.95$ the maximum 
of $m_{res}$ is reached for $N_s>28$). We can then speak of a sort of 
suppression of the chiral violations for small $N_s$, which becomes stronger
as $M$ approaches the borders,and it could be that in this case the density 
of eigenvalues or the radius of the modes in Eq.~(\ref{eq:transmat}) change 
in such a way to produce this kind of effect, or that this equation breaks down
in this region. It could also happen that the value of $N_s$ for which the
chiral violations reach their maximum (at fixed $M$) is further and further 
shifted toward higher values of $N_s$ until one cannot observe any exponential 
decay at all even for very large $N_s$, and possibly for all $N_s$, as in the 
exact asymptotic results of Eqs. (\ref{eq:limtad0}) and (\ref{eq:limtad1}).

In Tables \ref{tab:limitvafeynman} and \ref{tab:limitvalandau} we can observe a
similar behavior in the case of the difference of the vector and axial-vector 
currents. Since $c_{mix}$ apparently does not instead show these effects, it 
could be that this quantity is described by a somewhat different formula than 
Eq.~(\ref{eq:transmat}).

It would be interesting to carry out further studies about these phenomena
which happen near the borders of $M$, and more investigations in the future 
could clarify these issues. We have however seen that already at one loop we 
can observe interesting features, in some cases corresponding to what is 
expected from general considerations which can be derived from other methods.

\begin{table}[htp]
\caption{\label{tab:limitmresfeynman}
The residual mass near $M=2$, at $\beta = 5.2$ in Feynman gauge.} 
\begin{ruledtabular}    
\begin{tabular}{|c|rrrrrr|} 
   $M$ & $N_s=8$ &$N_s=12$ &$N_s=16$ &$N_s=20$ & $N_s=24$& $N_s=28$  
      \vspace{0.05cm} \\ \hline \vspace{-0.3cm} \\  
 1.91  &  0.043  &  0.094  &  0.120  &  0.127  &  0.120  &  0.107  \\
 1.92  &  0.034  &  0.085  &  0.115  &  0.128  &  0.128  &  0.120  \\
 1.93  &  0.026  &  0.074  &  0.106  &  0.125  &  0.132  &  0.130  \\
 1.94  &  0.018  &  0.061  &  0.094  &  0.117  &  0.130  &  0.134  \\
 1.95  &  0.011  &  0.048  &  0.080  &  0.104  &  0.121  &  0.132  \\ \hline
\end{tabular}
\end{ruledtabular}
\end{table}

\begin{table}[htp]
\caption{\label{tab:limitmreslandau}
The residual mass near $M=2$, at $\beta = 5.2$ in Landau gauge.} 
\begin{ruledtabular}    
\begin{tabular}{|c|rrrrrr|} 
   $M$ & $N_s=8$ &$N_s=12$ &$N_s=16$ &$N_s=20$ & $N_s=24$& $N_s=28$  
      \vspace{0.05cm} \\ \hline \vspace{-0.3cm} \\  
 1.91  &  0.049  &  0.099  &  0.124  &  0.130  &  0.122  &  0.108  \\
 1.92  &  0.040  &  0.089  &  0.119  &  0.131  &  0.130  &  0.122  \\
 1.93  &  0.031  &  0.078  &  0.110  &  0.128  &  0.134  &  0.132  \\
 1.94  &  0.023  &  0.065  &  0.098  &  0.120  &  0.132  &  0.137  \\
 1.95  &  0.015  &  0.052  &  0.083  &  0.107  &  0.124  &  0.134  \\ \hline
\end{tabular}
\end{ruledtabular}
\end{table}

\begin{table}[htp]
\caption{\label{tab:limitvafeynman}
Coefficient of $\bar g^2$ for $\Delta$ near $M=2$, in Feynman gauge.} 
\begin{ruledtabular}    
\begin{tabular}{|c|rrrrrr|} 
   $M$ & $N_s=8$ &$N_s=12$ &$N_s=16$ &$N_s=20$ & $N_s=24$& $N_s=28$  
      \vspace{0.05cm} \\ \hline \vspace{-0.3cm} \\  
 1.94  &  3.699  &  3.151  &  2.488  &  1.915  &  1.467  &  1.125  \\
 1.95  &  4.246  &  3.858  &  3.220  &  2.594  &  2.063  &  1.637  \\
 1.96  &  4.885  &  4.753  &  4.220  &  3.592  &  2.995  &  2.475  \\
 1.97  &  5.630  &  5.883  &  5.586  &  5.063  &  4.474  &  3.897  \\
 1.98  &  6.494  &  7.301  &  7.442  &  7.225  &  6.823  &  6.332  \\
 1.99  &  7.495  &  9.073  &  9.945  & 10.377  & 10.524  & 10.481  \\ \hline
\end{tabular}
\end{ruledtabular}
\end{table}

\begin{table}[htp]
\caption{\label{tab:limitvalandau}
Coefficient of $\bar g^2$ for $\Delta$ near $M=2$, in Landau gauge.} 
\begin{ruledtabular}    
\begin{tabular}{|c|rrrrrr|} 
   $M$ & $N_s=8$ &$N_s=12$ &$N_s=16$ &$N_s=20$ & $N_s=24$& $N_s=28$  
      \vspace{0.05cm} \\ \hline \vspace{-0.3cm} \\  
 1.94  &  2.112  &  1.897  &  1.558  &  1.237  &  0.973  &  0.763  \\
 1.95  &  2.416  &  2.310  &  1.997  &  1.654  &  1.346  &  1.090  \\
 1.96  &  2.774  &  2.836  &  2.602  &  2.268  &  1.929  &  1.621  \\
 1.97  &  3.193  &  3.503  &  3.432  &  3.179  &  2.856  &  2.522  \\
 1.98  &  3.682  &  4.346  &  4.564  &  4.522  &  4.335  &  4.071  \\
 1.99  &  4.251  &  5.402  &  6.098  &  6.488  &  6.673  &  6.718  \\ \hline
\end{tabular}
\end{ruledtabular}
\end{table}

\section{Conclusions}
\label{sec:concl}

In this article we have presented the calculation of a few one-loop 
amplitudes for domain-wall fermions at finite $N_s$, with the intention 
of gathering some knowledge about the extent of chiral symmetry breaking for 
choices of $N_s$ which are far away from the case of infinite extent in the 
extra fifth dimension. In particular, we have studied three quantities whose 
deviations from their values at $N_s=\infty$ can provide some significant 
estimates of chiral violations: the residual mass, the difference between 
the vector and axial-vector renormalization constants, and a power-divergent 
mixing of a deep-inelastic operator which is entirely due to the breaking
of chirality.

We have automated the perturbative calculations by developing suitable FORM 
codes. We have found that the pattern of the deviations from the case of exact 
chirality turns out to be approximately the same for all quantities studied, 
that is these violations substantially increase when $N_s$ becomes small or 
when $M$ approaches 0 or 2, i.e., close to the borders of the region of allowed
values of $M$. Our perturbative calculations show indeed that the numerical 
deviations from the case of infinite extension in the fifth dimension depend, 
apart from $N_s$ and to a smaller extent from the bare coupling $g_0$, very 
strongly on the choice of $M$. These deviations can become rather pronounced 
when $M$ is close to the borders of the region of allowed values. For $M=1.8$, 
a standard choice in Monte Carlo simulations, chiral violations remain still 
not small for $N_s=16$, as can be seen from $m_{res}$, which for a lattice
spacing of 2 GeV is equal to about 100 MeV in the quenched case and about 
120 MeV in full QCD. For the difference between the vector and axial-vector 
renormalization constants as well as for the power-divergent mixing the chiral 
violations are instead of about 2-3 MeV, suggesting that they are of higher 
order in $m_{res}$.

We think that to the extent that one-loop perturbation theory can provide hints
to the actual behavior of these quantities, and if higher-order corrections do 
not strongly compensate for these 1-loop effects, our work shows that chiral 
violations at $N_s=16$ are still somewhat rather pronounced, and it suggests 
that one should perhaps consider performing simulations which use at least 
$N_s=20$ or $N_s=24$ points in the fifth dimension, in order to be able to 
work with reasonably small chiral violations. 

Our work also confirms (though only from the numerical side) the legitimacy
of the assumptions made in previous calculations at infinite $N_s$,
where from the start it was postulated that the limit $N_s\to\infty$ 
for the quark propagators could be taken before doing the actual computations 
of the Feynman diagrams and the numerical integration. The asymptotic 
propagators introduce in fact spurious infrared singularities not present in 
the theory at finite $N_s$, and in passing from $N_s=\infty$ to any finite 
value of $N_s$ the analytical Atiyah-Singer index of the Dirac operator, 
which protects the quarks from acquiring a nonzero mass, disappears. 
One cannot then be certain that this transition occurs with a smooth and 
continuous behavior, although at the end it is likely that no problem 
is present \cite{Neuberger:1997bg}. 
We can observe that the results which we have presented in this work rapidly 
approach for large $N_s$ the numbers which were obtained with the simplified 
asymptotic theory, with the exception of the regions very close to the borders 
of allowed values, $M\to 0$ or $M\to 2$, where other effects can come into 
play. We can then a posteriori confirm that, at least for one-loop calculations
made at choices of $M$ which are not too extreme, the inversion of the limits 
was numerically legitimate.

A disturbing finding of our calculations is the pathological behavior of 
renormalization factors, which are no longer gauge invariant, although the 
deviations from gauge invariance are not numerically large, even for values 
of $M$ far away from 1 (the pattern of chiral violations seems to be the same
for all quantities considered in this work). It could be that the act of 
choosing a definite gauge affects the amount of chirality-breaking effects in 
Monte Carlo simulations as well. 
Furthermore, the coefficients of the logarithmic divergences turn out to 
depend on $N_s$ and $M$, and only for $N_s=\infty$ they are the same as 
the ones calculated in the continuum. In particular, the anomalous dimension
of the vector (as well as the axial-vector) current is not zero, and not
constant in $N_s$ and $M$. It would be interesting in the future to investigate
in more detail these phenomena, which could perhaps contribute to a fuller 
understanding of domain-wall simulations.

\begin{acknowledgments}
I am grateful for the support by Fonds zur F\"orderung der Wissenschaftlichen 
Forschung in \"Osterreich (FWF), Project P16310-N08. 
\end{acknowledgments}


\begin{thebibliography}{99}

\bibitem{Kaplan:1992bt}
  D.~B.~Kaplan,
  Phys.\ Lett.\ B {\bf 288}, 342 (1992)
  [arXiv:hep-lat/9206013].

\bibitem{Shamir:1993zy}
  Y.~Shamir,
  Nucl.\ Phys.\ B {\bf 406}, 90 (1993)
  [arXiv:hep-lat/9303005].

\bibitem{Furman:1994ky}
  V.~Furman and Y.~Shamir,
  Nucl.\ Phys.\ B {\bf 439}, 54 (1995)
  [arXiv:hep-lat/9405004].

\bibitem{Ginsparg:1981bj}
  P.~H.~Ginsparg and K.~G.~Wilson,
  Phys.\ Rev.\ D {\bf 25}, 2649 (1982).

\bibitem{Luscher:1998pq}
  M.~L\"uscher,
  Phys.\ Lett.\ B {\bf 428}, 342 (1998)
  [arXiv:hep-lat/9802011].

\bibitem{Yamada:1997bj}
  A.~Yamada,
  Phys.\ Rev.\ D {\bf 57}, 1433 (1998)
  [arXiv:hep-lat/9705040].

\bibitem{Yamada:1997hc}
  A.~Yamada,
  Nucl.\ Phys.\ B {\bf 514}, 399 (1998)
  [arXiv:hep-lat/9707032].

\bibitem{Kikukawa:2001mw}
  Y.~Kikukawa,
  Phys.\ Rev.\ D {\bf 65}, 074504 (2002)
  [arXiv:hep-lat/0105032].

\bibitem{Blum:1996jf}
  T.~Blum and A.~Soni,
  Phys.\ Rev.\ D {\bf 56}, 174 (1997)
  [arXiv:hep-lat/9611030].

\bibitem{Blum:1997mz}
  T.~Blum and A.~Soni,
  Phys.\ Rev.\ Lett.\  {\bf 79}, 3595 (1997)
  [arXiv:hep-lat/9706023].

\bibitem{Aoki:2000pc}
  S.~Aoki, T.~Izubuchi, Y.~Kuramashi and Y.~Taniguchi,
  Phys.\ Rev.\ D {\bf 62}, 094502 (2000)
  [arXiv:hep-lat/0004003].

\bibitem{Chen:2000zu}
  P.~Chen {\it et al.},
  Phys.\ Rev.\ D {\bf 64}, 014503 (2001)
  [arXiv:hep-lat/0006010].

\bibitem{Blum:2000kn}
  T.~Blum {\it et al.},
  Phys.\ Rev.\ D {\bf 69}, 074502 (2004)
  [arXiv:hep-lat/0007038].

\bibitem{Blum:2001xb}
  T.~Blum {\it et al.}  [RBC Collaboration],
  Phys.\ Rev.\ D {\bf 68}, 114506 (2003)
  [arXiv:hep-lat/0110075].

\bibitem{Noaki:2001un}
  J.~I.~Noaki {\it et al.}  [CP-PACS Collaboration],
  Phys.\ Rev.\ D {\bf 68}, 014501 (2003)
  [arXiv:hep-lat/0108013].

\bibitem{Aoki:2002vt}
  Y.~Aoki {\it et al.},
  Phys.\ Rev.\ D {\bf 69}, 074504 (2004)
  [arXiv:hep-lat/0211023].

\bibitem{Aoki:2004ht}
  Y.~Aoki {\it et al.},
  Phys.\ Rev.\ D {\bf 72}, 114505 (2005)
  [arXiv:hep-lat/0411006].

\bibitem{Orginos:2005uy}
  K.~Orginos, T.~Blum and S.~Ohta,
  Phys.\ Rev.\ D {\bf 73}, 094503 (2006)
  [arXiv:hep-lat/0505024].

\bibitem{Aoki:2005ga}
  Y.~Aoki {\it et al.},
  Phys.\ Rev.\ D {\bf 73}, 094507 (2006)
  [arXiv:hep-lat/0508011].

\bibitem{Negele:2005za}
  J.~W.~Negele {\it et al.}  [LHPC Collaboration],
  Int.\ J.\ Mod.\ Phys.\ A {\bf 21}, 720 (2006)
  [arXiv:hep-lat/0509101].

\bibitem{Yamada:2005dv}
  N.~Yamada, T.~Blum, M.~Hayakawa and T.~Izubuchi  [RBC Collaboration],
  PoS {\bf LAT2005}, 092 (2005)
  [arXiv:hep-lat/0509124].

\bibitem{Lin:2005gh}
  M.~f.~Lin,
  PoS {\bf LAT2005}, 094 (2005)
  [arXiv:hep-lat/0509178].

\bibitem{Edwards:2005kw}
  R.~G.~Edwards {\it et al.}  [LHPC Collaboration],
  PoS {\bf LAT2005}, 056 (2005)
  [arXiv:hep-lat/0509185].

\bibitem{Dawson:2005zv}
  C.~Dawson, T.~Izubuchi, T.~Kaneko, S.~Sasaki and A.~Soni,
  PoS {\bf LAT2005}, 337 (2005)
  [arXiv:hep-lat/0510018].

\bibitem{Noaki:2005zw}
  J.~Noaki,
  PoS {\bf LAT2005}, 350 (2005)
  [arXiv:hep-lat/0510019].

\bibitem{Edwards:2005ym}
  R.~G.~Edwards {\it et al.}  [LHPC Collaboration],
  Phys.\ Rev.\ Lett.\  {\bf 96}, 052001 (2006)
  [arXiv:hep-lat/0510062].

\bibitem{Ohta:2005cn}
  S.~Ohta, H.~Lin and N.~Yamada  [RBC Collaboration],
  PoS {\bf LAT2005}, 096 (2005)
  [arXiv:hep-lat/0510071].

\bibitem{Hashimoto:2005re}
  K.~Hashimoto, T.~Izubuchi and J.~Noaki  [RBC-UKQCD Collaborations],
  PoS {\bf LAT2005}, 093 (2005)
  [arXiv:hep-lat/0510079].

\bibitem{Berruto:2005hg}
  F.~Berruto, T.~Blum, K.~Orginos and A.~Soni,
  Phys.\ Rev.\ D {\bf 73}, 054509 (2006)
  [arXiv:hep-lat/0512004].

\bibitem{Berruto:2005sy}
  F.~Berruto, T.~Blum, K.~Orginos and A.~Soni,
  PoS {\bf LAT2005}, 010 (2005).

\bibitem{Lin:2006kg}
  H.~W.~Lin, S.~Ohta and N.~Yamada  [RBC Collaboration],
  Nucl.\ Phys.\ Proc.\ Suppl.\  {\bf 153}, 199 (2006).

\bibitem{Vranas:1997da}
  P.~M.~Vranas,
  Phys.\ Rev.\ D {\bf 57}, 1415 (1998)
  [arXiv:hep-lat/9705023].

\bibitem{Vranas:1997ib}
  P.~M.~Vranas,
  Nucl.\ Phys.\ Proc.\ Suppl.\  {\bf 63}, 605 (1998)
  [arXiv:hep-lat/9709119].

\bibitem{Jung:2000fh}
  C.~Jung, R.~G.~Edwards, X.~D.~Ji and V.~Gadiyak,
  Phys.\ Rev.\ D {\bf 63}, 054509 (2001)
  [arXiv:hep-lat/0007033].

\bibitem{Gadiyak:2002ig}
  V.~G.~Gadiyak,
  Ph.D. Thesis, University of Maryland, UMI-30-78308 (2002).

\bibitem{Edwards:2005an}
  R.~G.~Edwards, B.~Joo, A.~D.~Kennedy, K.~Orginos and U.~Wenger,
  PoS {\bf LAT2005}, 146 (2005)
  [arXiv:hep-lat/0510086].

\bibitem{Aoki:1997xg}
  S.~Aoki and Y.~Taniguchi,
  Phys.\ Rev.\ D {\bf 59}, 054510 (1999)
  [arXiv:hep-lat/9711004].

\bibitem{Aoki:1998vv}
  S.~Aoki, T.~Izubuchi, Y.~Kuramashi and Y.~Taniguchi,
  Phys.\ Rev.\ D {\bf 59}, 094505 (1999)
  [arXiv:hep-lat/9810020].

\bibitem{Blum:1999xi}
  T.~Blum, A.~Soni and M.~Wingate,
  Phys.\ Rev.\ D {\bf 60}, 114507 (1999)
  [arXiv:hep-lat/9902016].

\bibitem{Capitani:2005vb}
  S.~Capitani,
  Phys.\ Rev.\ D {\bf 73}, 014505 (2006)
  [arXiv:hep-lat/0510091].

\bibitem{Kikukawa:1997tf}
  Y.~Kikukawa, H.~Neuberger and A.~Yamada,
  Nucl.\ Phys.\ B {\bf 526}, 572 (1998)
  [arXiv:hep-lat/9712022].

\bibitem{Christ:2005xh}
  N.~Christ  [RBC and UKQCD Collaborations],
  PoS {\bf LAT2005}, 345 (2005).

\bibitem{Golterman:2003qe}
  M.~Golterman and Y.~Shamir,
  Phys.\ Rev.\ D {\bf 68}, 074501 (2003)
  [arXiv:hep-lat/0306002].

\bibitem{Golterman:2004cy}
  M.~Golterman, Y.~Shamir and B.~Svetitsky,
  Phys.\ Rev.\ D {\bf 71}, 071502 (2005)
  [arXiv:hep-lat/0407021].

\bibitem{Golterman:2005fe}
  M.~Golterman, Y.~Shamir and B.~Svetitsky,
  Phys.\ Rev.\ D {\bf 72}, 034501 (2005)
  [arXiv:hep-lat/0503037].

\bibitem{Svetitsky:2005qa}
  B.~Svetitsky, Y.~Shamir and M.~Golterman,
  PoS {\bf LAT2005}, 129 (2005)
  [arXiv:hep-lat/0508015].

\bibitem{Antonio:2005wm}
  D.~J.~Antonio {\it et al.}  [RBC and UKQCD Collaborations],
  PoS {\bf LAT2005}, 141 (2005).

\bibitem{Draper:2005mh}
  T.~Draper {\it et al.},
  PoS {\bf LAT2005}, 120 (2005)
  [arXiv:hep-lat/0510075].

\bibitem{Narayanan:1992wx}
  R.~Narayanan and H.~Neuberger,
  Phys.\ Lett.\ B {\bf 302}, 62 (1993)
  [arXiv:hep-lat/9212019].

\bibitem{Martinelli:1982mw}
  G.~Martinelli and Y.~C.~Zhang,
  Phys.\ Lett.\ B {\bf 123}, 433 (1983).

\bibitem{Capitani:2002mp}
  S.~Capitani,
  Phys.\ Rept.\  {\bf 382}, 113 (2003)
  [arXiv:hep-lat/0211036].

\bibitem{Neuberger:1997bg}
  H.~Neuberger,
  Phys.\ Rev.\ D {\bf 57}, 5417 (1998)
  [arXiv:hep-lat/9710089].

\bibitem{Vermaseren:2000nd}
  J.~A.~M.~Vermaseren,
  arXiv:math-ph/0010025.

\bibitem{Duerr} I acknowledge conversations with Stephan D\"urr about the
  scalar and pseudoscalar correlators.

\bibitem{Capitani:2000wi}
  S.~Capitani,
  Nucl.\ Phys.\ B {\bf 592}, 183 (2001)
  [arXiv:hep-lat/0005008].

\bibitem{Capitani:2000aq}
  S.~Capitani,
  Nucl.\ Phys.\ B {\bf 597}, 313 (2001)
  [arXiv:hep-lat/0009018].

\bibitem{Beccarini:1995iv}
  G.~Beccarini, M.~Bianchi, S.~Capitani and G.~Rossi,
  Nucl.\ Phys.\ B {\bf 456}, 271 (1995)
  [arXiv:hep-lat/9506021].

\bibitem{Gockeler:1995wg}
  M.~G\"ockeler, R.~Horsley, E.~M.~Ilgenfritz, H.~Perlt, P.~Rakow, G.~Schierholz and A.~Schiller,
  Phys.\ Rev.\ D {\bf 53}, 2317 (1996)
  [arXiv:hep-lat/9508004].

\end{thebibliography}
\end{document}